\documentclass[twoside,12pt]{article}
\usepackage{epsfig}
\usepackage{amsmath,amsfonts}

\newcommand{\be}{\begin{equation}}
\newcommand{\ee}{\end{equation}}
\newcommand{\bea}{\begin{eqnarray}}
\newcommand{\eea}{\end{eqnarray}}

\begin{document}

\title{R-mode constraints from neutron star equation of state  }

\author{M.C. Papazoglou and Ch.C. Moustakidis \\
$^{}$ Department of Theoretical Physics, Aristotle University of
Thessaloniki, \\ 54124 Thessaloniki, Greece }

\maketitle

\begin{abstract}
The gravitational radiation has been proposed a long time before, as an explanation for the observed relatively low spin frequencies of young neutron stars and of accreting neutron stars in low-mass X-ray binaries as well. In the present work we studied  the effects of the neutron star equation of  state  on the r-mode instability window  of rotating neutron stars. Firstly, we  employed  a set  of analytical solution of the Tolman-Oppemheimer-Volkoff equations  with special emphasis on the Tolman VII solution.  In particular, we tried to clarify the effects of the bulk neutron star properties (mass, radius, density distribution, crust size and elasticity) on the r-mode instability window. We found that the critical angular velocity $\Omega_c$ depends  mainly on the neutron star radius. The effects of the gravitational mass and the mass distribution are almost negligible. Secondly, we  studied  the effect of the elasticity of the crust, via to the slippage factor $S$ and also the effect of the nuclear equation of state, via the slope parameter $L$,  on the instability window. We found that the crust effects are more pronounced, compared to those originated from the equation of state. Moreover, we proposed simple analytical expressions which relate the macroscopic quantity $\Omega_c$ to the  radius, the parameter $L$ and the factor ${\cal S}$.  We also investigated   the possibility to measure the radius of a neutron star and the factor ${\cal S}$ with the help of accurate measures of $\Omega_c$ and the neutron star temperature. Finally, we studied the effects of the mutual friction on the instability window  and discussed the results in comparison with previous similar studies.
\\
\\
\end{abstract}

\section{Introduction}
There are several open problems in physics and astrophysics on neutron stars~\cite{Shapiro-83,Glendenning-2000,Haensel-07,Lattimer-012}. One of the problems  is why neutron stars do not spin up to the theoretically allowed limit called Kepler frequency. In particular, there is a sharp cut off for spins above 730, Hz which are well below the theoretically allowed upper limit~\cite{Prakash-014-a}. One possibility is the radiation of  gravitational waves from the rapidly rotating pulsars. In particular, neutron stars may suffer a number of instabilities which come in different flavors but they have a general feature in
common; they can be directly associated with unstable modes of
oscillation~\cite{Lidblom-2000a,Andersson-1998,Friedman-98,Friedman-99,Owen-98,Lidblom-98,Andersson-2001,
Andersson-2003,Kokkotas-99,Andersson-99,Bildsten-2000,Andersson-2000,Rieutord-2000,Bondarescu-09,Ho-011,Alford-2012b,Mahmoodifar-013, Gusakov-014,Kolomeitsev-014,Mytidis-015,Haskell-15,Kokkotas-015}.  The r-modes are oscillations of
rotating stars whose restoring force is the Coriolis force. The
gravitational radiation-driven instability of these modes has been
proposed as an explanation for the observed relatively low spin
frequencies of young neutron stars and of accreting neutron stars
in low-mass X-ray binaries as well \cite{Lidblom-2000a}. This
instability can only occur when the gravitational-radiation
driving time scale of the r-mode is shorter than the time scales
of the various dissipation mechanisms that may occur in the
interior of the neutron star.

The neutron star (NS) structure originates from the balance between the short-range character of the nuclear forces and the long-range gravitational field. In view of the above, NS is a unique laboratory  to  test various theories of  gravity and also to probe the nuclear equation of state both for low  and high baryonic densities.  This is one of the main reasons why neutron stars  are considered as the most exciting astrophysical objects. The density distribution of a neutron star is  defined by the hydrodynamic equilibrium as a result of  the interplay between the pressure of its ingredient particles and the gravity. Basically,   there are two ways to construct  the density distribution of a neutron star. The first one, is by solving numerically the Tolman-Oppemheimer-Volkoff (TOV) equations by employing a specific equation of state (EOS). This method  leads directly to a realistic density distribution  profile and also provide a configuration of pairs (M,R) with one to one correspondence  between them. The second one is to find analytical solution of the TOV equations by employing various density profiles. In this case, in  each pair $(M,R)$ corresponds an individual EOS. The analytical solutions have the significant  advantage of  being applicable for a wide range of pairs $(M,R)$ and desirably for a wide range of density distribution configurations. This is the main reason why the analytical solutions are suitable to study  relations that depend weakly on the EOS.

Actually, there are many  analytical solutions, however  nearly all of them are physically unrealistic. In particular, the known analytical solutions are  divided into two classes~\cite{Delgaty-98}. The first class is related to a neutron star in which the density $\rho$ and the pressure $P$ vanish at the surface. There are only three known analytical solutions that exhibit this behavior: the Tolman VII solution~\cite{Tolman-39}, the Buchdahl solution~\cite{Buchdahl-67} and the Naraiai solution~\cite{Nariai-50,Nariai-51,Nariai-99}. The second class of solutions is related to the called self-bounded stars, where while the pressure vanishes at the surface, the density remains finite. There is a huge number of solutions that exhibit this behavior but the useful ones are various versions of the Tolman IV~\cite{Delgaty-98} and VII~\cite{Delgaty-98,Postnikov-010} solutions and also the uniform density solution~\cite{Delgaty-98}. The above analytical solutions are valuable and interesting because one may study their properties in complete details. Also they are complementary to the numerical solutions~\cite{Adler-74}.

The motivation of the present work is twofold. First, we intend to examine possible constraints on the r-mode instability related to the bulk neutron stars properties (mass, radius, density distribution, crust elasticity,  e.t.c.) by employing a suitable set of analytical solutions of TOV equations. Most of the mentioned solutions have never been used for the study of the r-mode instabilities in neutron and quark stars
(the only exceptions are  the uniform solution, which due to its simplicity, it has been extensively used and the more realistic Tolman VII solution).
Since all of them are related directly both to the bulk neutron stars properties  as well as to  their stellar structure,
they are suitable for the study of the various instability modes (included r, f, w and etc.).  It is worth noting  that according to \cite{Andersson-2001} the use of realistic equation of state in a Newtonian framework may  imposes some uncertainties since there is  no one-to-one correspondence between Newtonian and relativistic stellar models. However, we consider that even in this case it is important to understand whether and in which extent the overall properties of the equation of state affect the r-mode instability.

Second, our aim is to examine and  if possible to establish,  relations between the critical angular velocity $\Omega_c$ and  a)  the nuclear equation of state via the slope parameter $L$ and b) the crust elasticity via the slippage factor ${\cal S}$. In particular, we propose a correlation between   $\Omega_c$ and the derivative of the nuclear symmetry energy with respect to the baryon density. This idea is based upon the empirical relation between the neutron star radius and the pressure of the neutron star matter for baryon densities, close and even higher to the saturation density,  which has been found by \cite{Lattimer-01}.  In any case, it will be of interest to inquire for possible connections between macrophysics properties of a NS (i.e. mass, radius, moment of inertia, critical angular velocity) and the microscopic ones (i.e the isovector character  of nuclear forces) and also to impose constraints on observation data from theoretical predictions and vice-versa.  Actually, there are some recent efforts to constrain the nuclear
physics input (for example the slope parameter $L$) by employing the
related observation data in low-mass X-ray binaries~\cite{Read-09-a,Read-09-b,Rezzolla-014,Chirenti-2015,Newton-014,Fattoyev-014,Watts-016}. In general, this is a very complex problem,
since the nuclear equation of state affects in different ways the
r-mode instability. Additional
work is needed as well, to illustrate further this point. The present work is dedicated to this effort.

The article is organized as followed. In Sec~II we briefly review the r-mode formalism. In Section III we present the connection between the nuclear equation of state and bulk neutron star properties. The results are presented and discussed in Sec. IV.  Sec. V summarizes the present study.

\section{R-mode instability formalism}

The {\it r}-modes evolve with time dependence $ e^{i \omega
t-t/\tau}$ as a consequence of ordinary hydrodynamics and the
influence of the various dissipative processes. The real part of
the frequency of these modes, $\omega$, is given by

\begin{equation}
\omega=-\frac{(l-1)(l+2)}{l+1}\Omega, \label{omega-1}
\end{equation}
where $\Omega$ is the angular velocity of the unperturbed star~\cite{Lidblom-98}.
The imaginary part $1/\tau$ is determined by the effects of
gravitational radiation, viscosity, etc.
\cite{Lidblom-2000a,Owen-98,Lidblom-98}. In the small-amplitude
limit, a mode is a driven, damped harmonic oscillator with an
exponential damping time scale
\begin{eqnarray}
\frac{1}{\tau(\Omega,T)}&=&\frac{1}{\tau_{_{GR}}(\Omega)}+
\frac{1}{\tau_{_{EL}}(\Omega,T)}+\frac{1}{\tau_{_{BV}}(\Omega,T)}\nonumber \\
&+&\frac{1}{\tau_{_{SV}}(\Omega,T)}+\frac{1}{\tau_{_{MF}}(\Omega,T)}\nonumber\\
&+&
{\rm additional \ terms},
\label{t-1}
\end{eqnarray}
where $\tau_{_{GR}}$, $\tau_{_{EL}}$, $\tau_{_{BV}}$,  $\tau_{_{SV}}$   and  $\tau_{_{MF}}$   are the gravitational
radiation time scale, the damping time scale due to viscous dissipation at the boundary layer of the rigid crust and fluid core, the  bulk and shear viscosity dissipation times scales respectively  and  the damping time scale due to the mutual friction.
Gravitational radiation tends to drive the {\it r}-modes unstable,
while viscosity  and mutual friction suppress  the instability. More precisely
dissipative effects cause the mode to decay exponentially as
$e^{-t/\tau}$ (i.e., the mode is stable) as long as $\tau >0$
\cite{Lidblom-98}.
In addition, the time scale $\tau$ is written
\begin{eqnarray}
\frac{1}{\tau(\Omega,T)}&=&\frac{1}{\tilde{\tau}_{_{GR}}}\left(\frac{\Omega}{\Omega_0}  \right)^{2l+2}+
\frac{1}{\tilde{\tau}_{_{SV}}}\left(\frac{10^9 K}{T}\right)^2\nonumber\\
&+&\frac{1}{\tilde{\tau}_{_{BV}}}\left(\frac{T}{10^9 K}\right)^6
\left(\frac{\Omega}{\Omega_0}  \right)^2\nonumber\\
&+&
\frac{1}{\tilde{\tau}_{_{EL}}}\left(\frac{10^8 K}{T}\right)
\left(\frac{\Omega}{\Omega_0}  \right)^{1/2} \nonumber\\
&+&\frac{1}{\tilde{\tau}_{_{MF}}}\left(\frac{\Omega}{\Omega_0}  \right)^{5} ,
\label{t-2}
\end{eqnarray}
where $\Omega_0=\sqrt{\pi G\overline{\rho}}$ and
$\overline{\rho}=3M/4\pi R^3$ is the mean density of the star. Moreover, the maximum angular velocity
$\Omega_{\rm K}$ (Kepler angular velocity) for any star occurs when the material at
the surface effectively orbits the star \cite{Lidblom-98}. This velocity is nearly
$\Omega_{{\rm K}}=\frac{2}{3}\Omega_0$

The damping time $\tau_i$ for the individual mechanisms is defined
in general by \cite{Lidblom-2000a}
\begin{equation}
\frac{1}{\tau_i}\equiv -\frac{1}{2E}\left(\frac{dE}{dt}\right)_i.
\label{tau-E-i}
\end{equation}
In Eq.~(\ref{tau-E-i})  the total energy $E$ of the r-mode is
given by~\cite{Lidblom-2000a,Lidblom-98}
\begin{equation}
E=\frac{1}{2}\alpha^2R^{-2l+2}\Omega^2 \int_0^R
\rho(r)r^{2l+2} dr, \label{E-1}
\end{equation}
where $\alpha$ is the dimensionless amplitude of the mode, $R$ is the radius,
$\Omega$ is the  angular velocity  and $\rho(r)$ is the radial dependence of the
mass density of the neutron star. In the present work we consider that the density distribution has the form $\rho(r)=\rho_c {\cal F}(x)$ where $\rho_c$ is the central density and ${\cal F}(x)$ is a dimensionless function of $x=r/R$.

The contribution of gravitational radiation to the imaginary part of the frequency of the mode $1/\tau_{_{GR}}$ is given by the expression
\cite{Lidblom-2000a,Lidblom-98}
\begin{eqnarray}
\frac{1}{\tau_{_{GR}}}&=&-\frac{32\pi G
\Omega^{2l+2}}{c^{2l+3}}\frac{(l-1)^{2l}}{[(2l+1)!!]^2}\left(\frac{l+2}{l+1}\right)^{2l+2}\nonumber \\
&\times&
\int_0^{R}\rho(r) r^{2l+2} dr \quad \left({\rm s}^{-1}\right). \label{tgr-1}
\end{eqnarray}
In various values of $l$ correspond different kinds of modes. In the present work we consider the smallest of these (the $l=2$ r-mode) and the time scale $\tau_{_{GR}}$ is written
\begin{eqnarray}
\frac{1}{\tau_{_{GR}}}&=&-7.661 \cdot 10^{-46}\left(\frac{\Omega}{{\rm Hz}}\right)^6 \left(\frac{R}{{\rm km}}  \right)^7 \nonumber \\
 &\times&\left(\frac{\rho_c}{{\rm gr \ cm^{-3}}}\right)
{\cal I}_1, \quad \left({\rm s}^{-1}\right),
\label{tau-GR-new-3}
\end{eqnarray}
where  the integral ${\cal I}_1$ is defined as
\begin{equation}
{\cal I}_1=\int_0^1{\cal F}(x) x^6 dx.
\label{str-int-1}
\end{equation}
The bulk viscosity is the main dissipation mechanism at high temperature. It originates from the variations of pressure and density due to the pulsation modes. This leads to the instability on the $\beta$-equilibrium in neutron star matter  and consequently to energy dissipation in order to recur the equilibrium~\cite{Vidana-012}. Actually, the treatment of  the viscosity  must be consistent with the employed equation of state. That is because the viscosity coefficients are functional of the equation of state. In this case,   different equation of states predict different viscosity coefficients. However, in the present work we treat the problem in an approximated way by using expressions which are extensively used in the literature, independently of the employed equation of state. In particular, the dissipation time scale  due to the bulk viscosity is given by \cite{Lidblom-98,Vidana-012}
\begin{eqnarray}
\frac{1}{\tau_{_{BV}}}&=&\frac{4\pi}{690}\left(\frac{\Omega}{\Omega_0} \right)^4R^{2l-2}\left(\int_0^R \rho(r) r^{2l+2} dr \right)^{-1}\nonumber\\
&\times&
\int_0^R \xi_{_{BV}} \left(\frac{r}{R}  \right)^6\left[ 1+0.86\left(\frac{r}{R}  \right)^2
\right]r^2 dr. \label{bulk-1}
\end{eqnarray}
The bulk viscosity $\xi_{_{BV}}$ for hot neutron star matter is given by \cite{Lidblom-98}
\begin{eqnarray}
\xi_{_{BV}}&=&6.0\times 10^{-59}\left(\frac{l+1}{2}\right)^2 \left(\frac{{\rm Hz}}{\Omega}  \right)^2  \nonumber\\
&\times&\left(\frac{\rho}{{\rm gr\ cm^{-3}}}\right)^{2} \left(\frac{T}{{\rm K}}\right)^{6} \quad ({\rm  gr \ cm^{-1}\  s^{-1}}).
 \label{bulk-2}
\end{eqnarray}
After some algebra we find
\begin{eqnarray}
\frac{1}{\tau_{_{BV}}}&=&0.248\cdot 10^{-91}\left(\frac{\Omega}{{\rm Hz}}\right)^2\left(\frac{\rho_c}{{\rm gr \ cm^{-3}}}\right)
\left(\frac{T}{{\rm K}}\right)^6 \nonumber \\
&\times& \left(\frac{M_{\odot}}{M}\right)^2\left(\frac{R}{{\rm km}} \right)^4
\frac{{\cal I}_2}{ {\cal I}_1} \quad ({\rm s^{-1}}),
\label{tau-BV-new-3}
\end{eqnarray}
where  the integral ${\cal I}_2$ is given by
\begin{equation}
{\cal I}_2=\int_0^1{\cal F}^2(x)x^8(1+0.86 x^2)dx.
\label{str-int-2}
\end{equation}

%
The shear viscosity is the dominant mechanism at low temperature and this is  due to  the momentum transport which takes place on the various kinds of particle-particle scattering. In particular there are several scattering processes which individually contribute to the total shear viscosity. In the present work we consider  two kinds of scattering processes: (a) the neutron-neutron scattering which is expected to be dominant   at $T>10^9$ K and  (b) the electron-electron scattering which is the main dissipation mechanism   at $T<10^9$. In general, the dissipation time scale  due to the shear viscosity is given by \cite{Lidblom-98}
\begin{eqnarray}
\frac{1}{\tau_{_{SV}}}&=&(l-1)(2l+1) \left(\int_0^R \rho(r) r^{2l+2} dr \right)^{-1}\nonumber\\
&\times&
\int_0^R \eta_{_{SV}} r^{2l} dr, \quad ({\rm s^{-1}}).
\label{shear-1}
\end{eqnarray}
The viscosity associated with the  neutron-neutron scattering and the electron-electron scattering are given respectively \cite{Lidblom-2000a,Lidblom-98,Kokkotas-99}
\begin{equation}
\eta_{nn}=347  \left(\frac{\rho}{{\rm gr\ cm^{-3}}}\right)^{9/4} \left(\frac{T}{{\rm K}}\right)^{-2}, \quad ({\rm g\ cm^{-1}\ s^{-1}}).
\label{eta-nn-1}
\end{equation}
\begin{equation}
\eta_{ee}=6.0\cdot 10^6 \left(\frac{\rho}{{\rm gr\ cm^{-3}}}\right)^2 \left(\frac{T}{{\rm K}}\right)^{-2}, \quad ({\rm g \ cm^{-1}}
\ s^{-1}). \label{eta-ee-1}
\end{equation}
The time scale $\tau_{_{SV}}^{nn}$ can also be written
 after some algebra
\begin{eqnarray}
\frac{1}{\tau_{_{SV}}^{nn}}&=&1.735 \cdot 10^{-7} \left(\frac{{\rm km}}{R} \right)^2\left(\frac{{\rm K}}{T} \right)^2
\nonumber\\
&\times&
\left(\frac{\rho_c}{{\rm gr \ cm^{-3}}}\right)^{5/4}
\frac{{\cal I}_3^{nn}}{I_1} \quad ({\rm s^{-1}}),
\label{tau-SV-new-2}
\end{eqnarray}
where ${\cal I}_3^{nn}$
\begin{equation}
{\cal I}_3^{nn}=\int_0^1{\cal F}^{9/4}(x) x^4 dx.
\label{str-int-3nn}
\end{equation}

The corresponding time scale $\tau_{SV}^{ee}$ is given by
\begin{equation}
\frac{1}{\tau_{SV}^{ee}}=3\cdot 10^{-3} \left(\frac{{\rm km}}{R} \right)^2\left(\frac{{\rm K}}{T} \right)^2
\left(\frac{\rho_c}{{\rm gr \ cm^{-3}}}\right)
\frac{{\cal I}_3^{ee}}{I_1} \quad ({\rm s^{-1}}),
\label{tau-SV-new-4}
\end{equation}
where ${\cal I}_3^{ee}$
\begin{equation}
{\cal I}_3^{ee}=\int_0^1{\cal F}^{2}(x) x^4 dx.
\label{str-int-3ee}
\end{equation}

Firstly, we study the case where the viscosity due to boundary layer of the rigid crust is not taken into account the equilibrium equation (minimal model). Then, the equilibrium equation, $\displaystyle \frac{1}{\tau}=0$,  is written
\begin{equation}
-\left(\frac{\Omega_c}{{\rm Hz}}\right)^6+a\left(\frac{\Omega_c}{{\rm Hz}}\right)^2+b=0.
\label{Omega-Eq-1}
\end{equation}
Eq.~(\ref{Omega-Eq-1}) is  directly converted  to a cubic equation. The above equation, in any case,  can be solved numerically to give the desired critical frequency  $\Omega_c$. However, in this case, it is  conceptually
difficult to intuit answers. Eq.~(\ref{Omega-Eq-1}) can be also  solved  analytically and the solution is given, for ${\cal Y} \leq 1 $, by
\begin{equation}
\Omega_c=\left(\frac{b}{2}\right)^{1/6}\sqrt{\left(1+\sqrt{1-{\cal Y}}  \right)^{1/3}+\left(1-\sqrt{1-{\cal Y}}  \right)^{1/3}}
\label{Omegac-T-all-1}
\end{equation}
and for ${\cal Y}\geq 1$ by
\begin{equation}
\Omega_c=\left(4b\sqrt{{\cal Y}}\right)^{1/6}\sqrt{\cos\left[\frac{1}{3}\tan^{-1}\left(\sqrt{{\cal Y}-1}  \right)  \right]}
\label{Omegac-T-all-2}
\end{equation}
where $\displaystyle {\cal Y}=\frac{4a^3}{27b^2}$ and also
\begin{equation}
a=3.237\cdot 10^4 \left(\frac{10 {\rm km}}{R} \right)^3\left(\frac{M}{M_{\odot}} \right)^2
\left(\frac{T}{10^9 {\rm K}} \right)^6\frac{{\cal I}_2}{{\cal I}_1^2},
\label{a-1}
\end{equation}
\begin{eqnarray}
b&=&2.265\cdot 10^{15}\left(\frac{10^9 {\rm K}}{T}  \right)^2 \left(\frac{10 {\rm km}}{R} \right)^9\frac{1}{{\cal I}_1^2}\nonumber\\
&\times&
\left[\left(\frac{\rho_c}{10^{16}{\rm gr \ cm^{-3}}}\right)^{1/4}{\cal I}_3^{nn}+1.729{\cal I}_3^{ee}    \right].
\label{b-1}
\end{eqnarray}
The analytical solutions (\ref{Omegac-T-all-1}) and (\ref{Omegac-T-all-2}) as far as we know, are displayed  for the first time in the literature.  These  solutions  provide some useful insights and can be easily elaborated  in order to lead in various relevant approximations.
In Eqs.~(\ref{Omegac-T-all-1}) and (\ref{Omegac-T-all-2}) is clearly  exhibited  the dependence of $\Omega_c$ on the bulk neutron stars structure properties $M$ and $R$, temperature $T$ as well as on the relative EOS via the structure integrals ${\cal I}_1$,${\cal I}_2$,${\cal I}_3^{nn}$ and ${\cal I}_3^{ee}$. It is worth pointing out that the mentioned expressions from the time scales and $\Omega_c$ are very general and  can be easily determined by employing analytical or/and numerical solution of the TOV equations.

We also consider   the effect on r-mode instability due to the presence of a solid crust in an old neutron star (minimal model+crust effects). It is proved that the presence of a viscous boundary layer under the solid crust of a neutron star increases the viscous damping rate of the fluid r-modes \cite{Lidblom-2000a,Bildsten-2000}. Actually, the presence of  a solid crust has a crucial effect on the r-mode motion and following the discussion of \cite{Andersson-2001} this effect can be understood as follows: based on the perfect fluid mode-calculations it is  anticipated the transverse motion associated with the mode at the crust-core
boundary to be large. However, if the crust is assumed to be rigid, the fluid motion must essentially fall off to zero at the base of the crust in order to satisfy a non-slip condition (in the rotating frame of reference).

When the dissipation effect, due to the crust, has been included the damping time scale  at the boundary
layer of the perfectly rigid crust and fluid core is given by
\cite{Lidblom-2000a}
\begin{eqnarray}
\tau_{_{EL}}&=&\frac{1}{2\Omega}\frac{2^{l+3/2}(l+1)!}{l(2l+1)!! {\cal
C}_l}\sqrt{\frac{2\Omega R_c^2\rho_{cr}}{\eta_{cr}}}\nonumber\\
&\times&
\int_0^{R_c}\frac{\rho(r)}{\rho_{cr}}\left(\frac{r}{R_c}\right)^{2l+2}
\frac{dr}{R_c} \quad ({\rm s}).
 \label{tv-1-mew}
\end{eqnarray}
The quantities $R_c$, $\rho_c$, $\eta_{cr}$ and $\rho_{cr}$ are the core radius,
the central density, the viscosity and density  of the fluid at the outer edge of the core respectively.
In deriving expression (\ref{tv-1-mew}) it is assumed that the crust
is rigid and hence static in the rotating frame. The motion of the
crust due to the mechanical coupling to the core effectively
increases $\tau_v$ by a factor of $(\Delta v/v)^{-2}$, where
$\Delta v/v$ denote the difference between the velocities in the
inner edge of the crust and the outer edge of the core divided by the
velocity of the core \cite{Levin-01}. Actually, the slippage factor ${\cal S}$ is defined as ${\cal S}=\Delta v/v$ (see the analysis below).
Considering the case $l=2$, where ${\cal C}_2=0.80411$ then the time scales originated from electron-electron and neutron-neutron scattering are given respectively
by
\begin{eqnarray}
\tau_{_{EL}^{ee}}&=&8.12\cdot 10 \left(\frac{{\rm Hz}}{\Omega} \right)^{1/2}\left(\frac{T}{{\rm K}}\right)
\left(\frac{{\rm gr \ cm^{-3}}}{\rho_{cr}}  \right)^{3/2}\nonumber\\
&\times&
\left(\frac{\rho_c}{\rm gr \ cm^{-3}} \right)\left(\frac{R}{{\rm km}} \right)^7
\left(\frac{{\rm km}}{R_c} \right)^6 \tilde{{\cal I}}_1 \quad ({\rm s})
 \label{tv-3-mew}
\end{eqnarray}
and
\begin{eqnarray}
\tau_{_{EL}^{nn}}&=&1.07\cdot 10^4\left(\frac{{\rm Hz}}{\Omega} \right)^{1/2}\left(\frac{T}{{\rm K}}\right)
\left(\frac{{\rm gr \ cm^{-3}}}{\rho_{cr}}  \right)^{13/8} \nonumber\\
&\times&
\left(\frac{\rho_c}{\rm gr \ cm^{-3}} \right)\left(\frac{R}{{\rm km}} \right)^7
\left(\frac{{\rm km}}{R_c} \right)^6 \tilde{{\cal I}}_1 \quad ({\rm s}),
 \label{tv-4-mew}
\end{eqnarray}
where
\begin{equation}
\tilde{{\cal I}}_1=\int_0^{x_c}{\cal F}(x)x^6 dx, \quad x_c=\frac{R_c}{R}
\label{I1-tilde}
\end{equation}
The gravitational radiation time scale is given by now
\begin{eqnarray}
\frac{1}{\tau_{_{GR}}}&=&-7.661 \cdot 10^{-46}\left(\frac{\Omega}{{\rm s}^{-1}}\right)^6 \left(\frac{R}{{\rm km}}  \right)^7 \nonumber\\
&\times&\left(\frac{\rho_c}{{\rm gr \ cm^{-3}}}\right)
\tilde{{\cal I}}_1 \quad ({\rm s^{-1}}).
\label{tau-GR-new-2-cr}
\end{eqnarray}
The bulk and shear viscosity time scales will be given again by Eqs.~(\ref{tau-BV-new-3}), (\ref{tau-SV-new-2}) and (\ref{tau-SV-new-4}) where now
the upper limit of the related integrals must be taken as $x_c=R_c/R$. In this case the relative integrals are defined as $\tilde{{\cal I}}_1$, $\tilde{{\cal I}}_2$, $\tilde{{\cal I}}_3^{ee}$ and $\tilde{{\cal I}}_3^{nn}$.

The equilibrium equation,  when the dissipation mechanism due to the crust has been included,  is given now by
\begin{equation}
-\left(\frac{\Omega_c}{{\rm Hz}}\right)^6+\tilde{a}\left(\frac{\Omega_c}{{\rm Hz}}\right)^2+\tilde{d}\left(\frac{\Omega_c}{{\rm Hz}}\right)^{1/2}+\tilde{b}=0,
\label{Omegac-crust-1}
\end{equation}
where the coefficients $\tilde{a}$  and $\tilde{b}$ are similar  with  $a$ and $b$, given by Eqs~(\ref{a-1}) and (\ref{b-1}), where now the structure integrals
${\cal I}_i$ ($ i=1,2,3$) have been replaced by the corresponding  $\tilde{{\cal I}}_i$. The coefficient $\tilde{d}$
is given by
\begin{eqnarray}
\tilde{d}&=&1.22\cdot 10^{24}\left(\frac{10^9 {\rm K}}{T}  \right)  \left(\frac{10 {\rm Km}}{R}\right)^{14} \left(\frac{R_c}{10 {\rm Km}}\right)^6
\nonumber\\
&\times&
\left(\frac{\rho_{cr}}{\rm gr \ cm^{-3}} \right)^{13/8}\left(\frac{\rm gr \ cm^{-3}}{\rho_c} \right)^2\frac{1}{\tilde{\cal I}_1^2}\nonumber\\
&\times&
\left(1+95.08\left(\frac{\rho_{cr}}{\rm gr \ cm^{-3}} \right)^{-1/8}    \right).
\label{tilde-d-1}
\end{eqnarray}
The density $\rho_{cr}$ which corresponds to the crust-core interface and its value is model dependent. In particular, the value of $\rho_{cr}$ is related to the phase transition between nuclei and uniform nuclear matter which takes place in the interior of a  neutron star and characterize the separate between the  solid crust and the fluid core. In this study  is taken to be $\rho_{cr}=1.5 \cdot 10^{14} \ {\rm g \ cm ^{-3}}$ which is very close to the theoretical estimation and the same values  used in similar studies~\cite{Lidblom-2000a}. In addition, the core radius $R_c$ is easily calculated by solving the equation $\rho(r)=\rho_{cr}$.

In the present work we also  explore  the case of an elastic crust. In this case the r-mode penetrates the crust and consequently the relative motion (slippage) between the crust and the core  is strongly reduced  compared to the rigid crust limit \cite{Levin-01}. In particular, the way the slippage factor ${\cal S}$ defined as ${\cal S}=\Delta v/v$ has been included on the r-mode problem which has been discussed in Refs.~\cite{Levin-01,Kinney-03,Glampedakis-06}. They propose that the factor $S$ must be included quadratically in the r-mode damping formula. This leads to a revised Ekman layer time scale~\cite{Glampedakis-06}
\begin{equation}
\tau_{_{EL}}^{\cal S}\rightarrow \frac{\tau_{_{EL}}}{{\cal S}^2}.
\label{S-1}
\end{equation}
Actually, the factor ${\cal S}$ depends mainly on the angular velocity $\Omega$, the core radius $R_c$ and the shear modulus $\mu$ but can be treated also, in  approximated way, as a constant (see also \cite{Glampedakis-06}). In particular, in Eq.~(\ref{S-1}) the factor ${\cal S}$ is  used as a free parameter varied in the interval of very low values (${\cal S}=0.05$) up to the value ${\cal S}=1$ which corresponds to a complete rigid crust. The effects of the factor ${\cal S}$ on $\Omega_c-T$ dependence are analyzed and discussed   in   Sec.~5.

Finally, in the present study we also consider  an additional damping mechanism called mutual friction (for more details see \cite{Lidblom-2000b} and \cite{Haskell-09}). This mechanism arises from the scattering of electrons of the magnetic fields which entrapped in the cores of the superfluid neutron vortices (\cite{Lidblom-2000b}). Mutual friction is considered as a candidate to provide the needed stability for the r-modes in old cold neutron stars while it has been shown that suppresses the gravitational radiation in the case of the f-modes of rotating neutron star. The dissipation time scale due to the mutual friction is given also by
\begin{equation}
\frac{1}{\tau_{_{MF}}}=3.2\cdot 10^{-28}\frac{1}{\tilde{\tau}_{_{MF}}}
\left(\frac{R}{{\rm km}}  \right)^{15/2}\left(\frac{M_{\odot}}{M}  \right)^{5/2}
\left(\frac{\Omega}{{\rm Hz}}  \right)^{5}.
\label{ta-MF-1}
\end{equation}
The characteristic damping time scale $\tilde{\tau}_{_{MF}}$ is independent of angular velocity and temperature (to lowest order) but sensitively depends on the entrainment parameter $\epsilon$ \cite{Lidblom-2000b}. Actually, $\tilde{\tau}_{_{MF}}$ has typical values $10^4$ sec, however,  a resonance phenomenon leads to very small values for  a few narrow range of $\epsilon$ (\cite{Lidblom-2000b}). In the present study we treat  $\tilde{\tau}_{_{MF}}$ as a phenomenological parameter varying in the range $5 \ {\rm s} \leq  \tilde{\tau}_{_{MF}} \leq 10^4\ {\rm s} $ according to the previous study of \cite{Lidblom-2000b}.

Now, the equilibrium equation  is given  by
\begin{equation}
\frac{\tilde{c}}{\tilde{\tau}_{_{MF}}}
\left(\frac{\Omega_c}{{\rm Hz}}  \right)^{5}    +\tilde{a}\left(\frac{\Omega_c}{{\rm Hz}}\right)^2+\tilde{d}\left(\frac{\Omega_c}{{\rm Hz}}\right)^{1/2}+\tilde{b}=\left(\frac{\Omega_c}{{\rm Hz}}\right)^6,
\label{Omegac-crust-MF-1}
\end{equation}
where  the coefficients $\tilde{a}$, $\tilde{b}$ and  $\tilde{d}$ is similar with those in Eq.~(\ref{Omegac-crust-1}) while the coefficient $\tilde{c}$
is given by
\begin{equation}
\tilde{c}=4.178\cdot 10^{17}\left(\frac{R}{{\rm km}}  \right)^{1/2}\left(\frac{M_{\odot}}{M}  \right)^{5/2}\left(\frac{{\rm gr \ cm^{-3}}}{\rho_c}\right)
\frac{1}{\tilde{I}_1}.
\label{tilde-c}
\end{equation}
%

\section{Nuclear equation of state relative to  r-mode studies }
Motivated by the strong radius dependence of the critical angular velocity $\Omega_c$, we propose a phenomenological approach to study the EOS effects on the r-mode instability window. This approach, despite its simplicity, provides a few insights of the mentioned study,  in a universal way,  and  also leads to some simplified empirical  relations. Moreover, the proposed method suggests and provides, in a way,  constraints on the nuclear equation of state with the help of accurate measurements of the main bulk neutron star properties.

We consider  that the energy per particle of nuclear matter close to saturation density $n_s$, in the parabolic approximation,  has the form~\cite{Lattimer-07}
\begin{equation}
 E(n,x)\simeq E(n,x=\frac{1}{2})+E_{sym}(n)(1-2x)^2.
\label{EOS-1}
\end{equation}
In Eq.~(\ref{EOS-1}) $n$ is the baryons  density,  $E_{sym}(n)$ is the symmetry energy and $x$ is the proton fraction.
$E(n,x=\frac{1}{2})$ is the energy per particle of symmetric nuclear matter, where close to the saturation density can be written in a good approximation
\begin{equation}
E(n,x=\frac{1}{2})\simeq -16 +\frac{{\cal K}}{18}\left(1-\frac{n}{n_s}\right)^2+\frac{{\cal L}}{162}\left(1-\frac{n}{n_s}\right)^3.
\label{Esym-1}
\end{equation}
The incompressibility  ${\cal K}$ and the skewness ${\cal L}$ are defined as
\begin{equation}
{\cal K}=9n_s^2\frac{\partial^2 E(n,x)}{\partial n^2}\mid_{n=n_s}
\label{K-1}
\end{equation}
and
\begin{equation}
{\cal L}=-27n_s^3\frac{\partial^3 E(n,x)}{\partial n^3}\mid_{n=n_s}.
\label{L-1}
\end{equation}
In neutron star matter, in order to satisfied  the  $\beta$-equilibrium, a small electron fraction exists and contributes to the total energy  according to
the expression
\begin{equation}
E_e=\frac{3 \hbar c}{4}(3\pi^2nx^4)^{1/3}.
\label{E-ele}
\end{equation}
The total energy is given now by
\begin{equation}
{\cal E}(n,x)=E(n,x)+E_e(n,x),
\label{EOS-2}
\end{equation}
while the  total pressure is defined as
\begin{equation}
P(n,x)=n^2\frac{\partial {\cal E}}{\partial n}.
\label{EOS-3}
\end{equation}
The proton fraction $x$ in $\beta$-equilibrium  is regulated by the value of the symmetry energy. In particular, is determined by solving the equation $\partial {\cal E}/\partial x=0$ which leads to~\cite{Prakash-94}
\begin{equation}
4E_{sym}(n)(1-2x)=\hbar c(3\pi^2 n x)^{1/3}.
\label{x-1}
\end{equation}
The combination of   Eqs.~(\ref{EOS-2}) and (\ref{EOS-3}) leads to
\begin{eqnarray}
 P(n,x)&=&n^2\left[\frac{\partial E_{sym}(n)}{\partial n}(1-2x)^2+\frac{xE_{sym}}{n}(1-2x)\right. \nonumber\\
 &-& \left.
\frac{{\cal K}}{9n_s}\left(1-\frac{n}{n_s}\right)-\frac{{\cal L}}{54 n_s}\left(1-\frac{n}{n_s}\right)^2  \right].
\label{EOS-4}
\end{eqnarray}
The expression (\ref{EOS-4}) has been extensively used in the literature for neutron star structure studies. In particular,
the pressure at the saturation density $n_s$ takes the form
\begin{eqnarray}
P(n_s,x_s)&=&n_s^2\left[\left(\frac{\partial E_{sym}(n)}{\partial n}\right)_{n_s}(1-2x_s)^2\right.\nonumber\\
&+&\left.\frac{x_sE_{sym}(n_s)}{n_s}(1-2x_s) \right].
\label{EOS-5}
\end{eqnarray}
Even more, close to the saturation density $n\simeq n_s$, and considering  that $n_s=0.16\ {\rm fm }^{-3}$ the proton fraction is small and  to a good approximation is given by
\begin{equation}
x_s\simeq \left(\beta+6\right)^{-1},
\label{xs-3}
\end{equation}
where
\begin{equation}
\beta=21.065\left(\frac{E_{sym}(n_s)}{30}\right)^{-3}.
\label{beta-2}
\end{equation}
Now, if we define the value of the symmetry energy at the saturation as $J=E_{sym}(n_s)$ and the slope parameter as  $L=3n_s\left(\frac{\partial E_{sym}(n)}{\partial n}\right)_{n_s}$,  Eq.~(\ref{EOS-5}) is rewritten as
\begin{equation}
P(n_s,x_s)=n_s\left[\frac{L}{3}(1-2x_s)^2+x_sJ(1-2x_s) \right].
\label{EOS-6}
\end{equation}
According to Eq.~(\ref{EOS-6}) the total pressure $P$ at the saturation density  depends directly  on the slope parameter $L$  (mainly) and
$J$ and indirectly on the mentioned parameters via the proton fraction $x_s$. Since the proton fraction, for densities close to $n_s$ is $x\ll 1$ then in a good approximation  Eq.~(\ref{EOS-6}) takes the form
\begin{equation}
P(n_s,x_s)\simeq n_s\frac{L}{3}.
\label{EOS-7}
\end{equation}
The expression (\ref{EOS-7}) has a clear meaning, the pressure of neutron star matter close to the saturation density is directly related to the symmetry energy via the slope parameter $L$. The above finding became very important when  Lattimer and Prakash, found a remarkable empirical relation which exists between the radii of $1$ and $1.4$ $M_{\odot}$ neutron stars and the corresponding neutron stars matter's pressure evaluated at densities $1$, $1.5$ and $2$ of the saturation density $n_s$~\cite{Lattimer-01}. The mentioned relation obeys a power-low relation:
\begin{equation}
R(M)= C(n,M) \left[\frac{P(n)}{{\rm MeV \ fm}^{-3}}\right]^{1/4},
\label{R-P-1}
\end{equation}
where  $R(M)$ is the radius of a star mass $M$, $P(n)$ is the pressure of neutron star matter at density $n$ and  $ C(n,M) $ is a number that depends on the density $n$ at which the pressure was evaluated and the stellar mass $M$. The values of  $ C(M,n) $ for the various cases are presented in Table.~3 of Ref.~\cite{Lattimer-01}. These values were estimated by averaging results of 31 disparate equations of state. Recently, Lattimer and Lim~\cite{Lattimer-013} excluding those equations of state, because  of  the maximum mass constraints imposed by PSR J1614-2230 (\cite{Demorest-010}) and  they  found the revised value
\begin{equation}
C(n_s,1.4M_{\odot})=9.52\pm 0.49 \ {\rm km}.
\label{CnM,1}
\end{equation}
The correlation (\ref{R-P-1}) is significant since the pressure of neutron star matter near the saturation density is, in large part, determined by the symmetry energy of the EOS \cite{Lattimer-01}. Moreover, it   relates the macroscopic  quantity $R$ (and of course all the relative quantities for example moment of inertia etc.) to the  microscopic quantity  $P$. Consequently, this formula, supports the statement that the nuclear equation of state plays an important role  on the construction of relativistic very dense objects i.t. a neutron star. Moreover the formula (\ref{R-P-1}), since it  directly relates the radius to the slope parameter $L$, exhibits the dependence of the neutron star size on the nuclear symmetry and  consequently on the isovector character of the nucleon-nucleon interaction.
More precisely, inverting equation (\ref{R-P-1}) yield
\begin{equation}
P(n)\simeq \left[\frac{R}{C(n,M)}  \right]^4 \ \left({\rm MeV \ fm}^{-3}   \right),
\label{R-P-3}
\end{equation}
where apparently, various restrictions on the equation of state are possible if the radius of a neutron star can be measured with high accuracy \cite{Lattimer-01}. As we show  in Sec.~5 the r-mode instability window, defined by the dependence $\Omega_c-T$, is strongly affected by the neutron star  radius $R$. The effects of the mass $M$ and the mass distribution $\rho(r)$ play minor role. Consequently, the dominant effect of the equation of state on the r-mode  is originated from the predicted values of the neutron star size. In view of the above statement,  we employ the correlation (\ref{R-P-1}) in order to relate  the angular velocity $\Omega_c$ with  effects of the EOS and mainly  the slope parameter $L$ which consists a basic characteristic of the EOS  and  is related to the derivative of the symmetry energy at the saturation density.

\section{Analytical solutions of the TOV equations }
For a static spherical symmetric system, the metric can be written as follows~\cite{Shapiro-83,Glendenning-2000}
\begin{equation}
ds^2=e^{\nu(r)}dt^2-e^{\lambda(r)}dr^2-r^2\left(d\theta^2+\sin^2\theta d\phi^2\right).
\label{GRE-1}
\end{equation}
The  density distribution and the local pressure related to the  metric functions $\lambda(r)$ and $\nu(r)$  according to the relations~\cite{Shapiro-83,Glendenning-2000}
\begin{equation}
\frac{8\pi G}{c^2}\rho(r)=\frac{1}{r^2}\left(1-e^{-\lambda(r)}\right)+ e^{-\lambda(r)}\frac{\lambda'(r)}{r},
\label{GRE-2}
\end{equation}
\begin{equation}
\frac{8\pi G}{c^4}P(r)=-\frac{1}{r^2}\left(1-e^{-\lambda(r)}\right)+ e^{-\lambda(r)}\frac{\nu'(r)}{r},
\label{GRE-3}
\end{equation}
where derivatives with respect to the radius are denoted by $'$.
The combination of Eqs.~(\ref{GRE-2}) and  (\ref{GRE-3})  leads to the  well known
Tolman-Oppenheimer-Volkoff equations~\cite{Shapiro-83,Glendenning-2000}
\begin{eqnarray}
\frac{dP(r)}{dr}&=&-\frac{G\rho(r) M(r)}{r^2}\left(1+\frac{P(r)}{\rho(r) c^2}\right)\nonumber\\
&\times&\left(1+\frac{4\pi P(r) r^3}{M(r)c^2}\right) \left(1-\frac{2GM(r)}{c^2r}\right)^{-1},
\label{TOV-1}
\end{eqnarray}
\begin{equation}
\frac{dM(r)}{dr}=4\pi r^2\rho(r).
\label{TOV-2}
\end{equation}
It is difficult to obtain exact solution of TOV equations in closed analytical form and they solved numerically with an equation of state specified. Actually, there are hundreds of analytical solutions of TOV equations but  three of them  satisfy the criteria that the pressure and energy density vanish on the surface of the star. Also  both of them  decrease monotonically with increasing radius. These three solutions, the Tolman VII, the  Buchdahl's and the Nariai IV are summarized below. Actually, the Tolman VII and  the  Buchdahl's have already be analyzed and employed in Ref.~\cite{Lattimer-01}.   However, since  the Nariai IV solution  is overlooked in the literature,  it is presented here more detailed (see also \cite{Lattimer-05}). It is worth  pointing out that all the analytical solutions presented and used in the present work contain two parameters,  the central density $\rho_c$ and the compactness parameter $\beta=GM/Rc^2$. All the mentioned solutions have been presented and analyzed with details in \cite{Postnikov-010,Lattimer-01,Lattimer-05,Lattimer-2000}.

\subsubsection*{Tolman VII solution}

The density distribution  is given by the simple analytical function~\cite{Tolman-39}
\begin{equation}
\rho(r)=\rho_c\left[1-\left(\frac{r}{R}\right)^2\right],\quad \rho_c=\frac{15M}{8\pi R^3},
\label{rho-TolVII}
\end{equation}
where obviously ${\cal F}(x)=1-x^2$. The core radius $R_c$ is given by the analytical expression
\begin{equation}
R_c=R\sqrt{1-1.263\cdot 10^{-4} \left(\frac{M_{\odot}}{M}  \right) \left(\frac{R}{{\rm km}}  \right)^3}.
\label{Rc-tovII}
\end{equation}

The central pressure becomes infinite for   $\beta > 0.3862$ and the causality is ensured if $\beta < 0.2698$. It is well known that despite its simplicity, this density distribution reproduces in a very good accuracy various neutron star properties including binding energy and moment of inertia while is in good agreement with realistic equation of state for neutron stars with $M >1 M_{\odot}$~\cite{Lattimer-01}. Moreover, the Tolman VII solution has the correct behavior not only on the extreme limits $r=0$ and $r=R$ but also in the intermediate regions (see Fig.~5 of \cite{Lattimer-01}). In addition, this solution has the interesting property that  for a given central density $\rho_c$ it has the greater maximum neutron star mass $M_{{\rm max}}$ and consequently sets an upper bound on $\rho_c$ for any measured neutron star mass \cite{Lattimer-05b}.

Additional,   the Tolman VII solution exhibits a density profile similar to the  density  profiles of  polytropic equations of state (solution of the Lane-Embden differential equation). All these polytropic density profiles  have a distinctive density falloff from the center to the edge of the Newtonian star and this is an expected feature of physical solutions. Recently, \cite{Raghoonundun-015} shown that the Tolman VII solution exhibits a polytropic behavior. They proved that this solution is at least as good as the Newtonian neutron stars, however with relativity being taken into account. In view of the above comments,  we consider that the Tolman VII solution is a very good approximation, since in a way, is a {\it bridge} to combine the Newtonian treatment of the r-mode instability  in a  relativistic star (neutron star).

\subsubsection*{Buchdahl solution}
The density distribution has the form~\cite{Buchdahl-67,Lattimer-01}
\begin{equation}
\rho=12\sqrt{P^*P}-5P,
\label{Buhd-den}
\end{equation}
where $P$ is the local pressure and $P^{*}$ is a parameter. While Buchdahl's solution has no particular physical basis, it does have two specific properties: (i) it can be made casual everywhere in the star by demanding that the local speed of sound $(dP/d\rho)^{1/2}$ be less than one  and   (ii) for small values of the pressure $P$ it reduces to $\rho=12\sqrt{P^*P}$,  which, in the Newtonian theory of stellar structure is the well known $n=1$ polytrope~\cite{Schutz-85}. So, Buchdahl's solution may be regarded as its relativistic generalization. The density distribution  can be expressed also as follows
\begin{equation}
\rho(r')=\frac{A^2uc^2}{4\pi G}(1-2\beta)(1-\beta-3u/2)(1-\beta+u)^{-2},
\label{Buhd-2}
\end{equation}
where $r'$, $u$, are radial-like variables defined as
\begin{eqnarray}
&&u=\beta \frac{\sin Ar'}{Ar'}, \quad r'=r(1-\beta+u)^{-1}(1-2\beta), \nonumber \\
&& A^2=288\pi P^{*}Gc^{-4}(1-2\beta)^{-1}.
\label{Buhd-3}
\end{eqnarray}
It is more convenient to use the variable $x'=r'/R$  instead of   $x=r/R$. The structure function is given now by
\begin{equation}
{\cal F}(x')=\frac{u}{\beta(1-5\beta/2)}(1-\beta-3u/2)(1-\beta+u)^{-2},
\label{Buhd-fx-1}
\end{equation}
where the variable $x'$ is defined in the interval
\begin{equation}
 0\leq x' \leq \frac{1-2\beta}{1-\beta}.
\label{Buhd-dx}
\end{equation}
Finally the central  density is given by
\begin{equation}
\rho_c=\frac{\pi M}{4R^3}\frac{\left(1-5\beta/2\right)(1-\beta)^2}{(1-2\beta)}.
\label{Buhd-rhoc}
\end{equation}
It is worth pointing out the limited domains of the Buchdals solutions. More precisely the conditions $\rho >0$, $c_s^2 >0$ and $c_s^2<c^2$ imply that $\beta<2/5$, $\beta<1/5$ and $\beta < 1/6$ correspondingly \cite{Lattimer-01}.

\subsubsection*{ Nariai IV solution}
The Nariai IV solution~\cite{Nariai-50,Nariai-51,Nariai-99}  is more complicated, compared to the previous ones,  and is used less in the literature. The analytical presentation here is based on the detailed analysis of \cite{Lattimer-05}.
The density distribution $\rho(r')$ is expressed in terms of  the parametric variable $r'$
\begin{eqnarray}
\frac{G}{c^2}\rho(r')&=&\frac{\sqrt{3\beta}}{4\pi R'^2(1-2\beta)}\frac{C^2}{E^2}\left[3\sin \tilde{f}(r')\cos \tilde{f}(r') \right.\nonumber\\ &-&\left.\sqrt{\frac{3\beta}{4}}\left(\frac{r'}{R'}\right)^2(3-\cos^2\tilde{f}(r')
\right],
\label{Nariai-1}
\end{eqnarray}
where
\begin{equation}
r=\frac{E}{C}\frac{r'}{\cos \tilde{f}(r')}\sqrt{1-2\beta}, \quad R'=\frac{R C}{\sqrt{1-2\beta}},
\label{Nariai-4}
\end{equation}
and
\begin{eqnarray}
\tilde{f}(r')&=&\cos^{-1}E+\sqrt{\frac{3\beta}{4}}\left[1-\left(\frac{r'}{R'} \right)^2\right], \nonumber \\
\tilde{g}(r')&=&\cos^{-1}C+\sqrt{\frac{3\beta}{2}}\left[1-\left(\frac{r'}{R'} \right)^2\right],
\label{Nariai-1-1}
\end{eqnarray}
\begin{eqnarray}
E^2&=&\cos^2\tilde{f}(R')=\frac{2+\beta+2\sqrt{1-2\beta}}{4+\beta/3},  \\
C^2&=&\cos^2\tilde{g}(R')\nonumber\\
&=&\frac{2E^2}{2E^2+(1-E^2)(7E^2-3)^2(5E^2-3)^{-2}}.\nonumber
\label{Nariai-2}
\end{eqnarray}
Now the density distribution can be written as
\begin{eqnarray}
\rho(r')&=&\frac{\sqrt{3}}{4\pi\sqrt{\beta}E^2}\left(\frac{c^2\beta}{G}\right)^3\frac{1}{M^2}
\left[3\sin \tilde{f}(r')\cos \tilde{f}(r')\right. \nonumber\\
&-&\left.\sqrt{\frac{3\beta}{4}}\left(\frac{r'}{R'}\right)^2(3-\cos^2\tilde{f}(r'))
\right].
\label{Nariai-1-b}
\end{eqnarray}
The central value of the density $\rho_c=\rho(r'=0)$ is given by the expression
\begin{equation}
\rho_c=\frac{3M}{8\pi R^3}\left[\left(\alpha-1 \right)\cos\sqrt{3\beta}
+\frac{6-\alpha}{\sqrt{3\beta}}\sin\sqrt{3\beta}\right],
\label{rhoc-narai-b}
\end{equation}
where $\alpha=\frac{3}{E^2}$. It is more convenient now to use the variable $x'=r'/R$  instant of the variable $x=r/R$  where
\begin{equation}
0\leq x'\leq \frac{R'}{R}=\frac{C}{\sqrt{1-2\beta}}.
\label{fx-trans}
\end{equation}
The distribution function ${\cal F}(x')$ can be written now
\begin{eqnarray}
{\cal F}(x')&=&\frac{2}{\sqrt{3\beta} E^2}
\left[3\sin \tilde{f}(x')\cos \tilde{f}(x') \right.\nonumber \\
&-&\left. \sqrt{\frac{3\beta}{4}} \left(x'\frac{R}{R'}\right)^2(3-\cos^2\tilde{f}(x'))
\right] \\
&\times&
\left[\left(\alpha-1 \right)\cos\sqrt{3\beta}
+\frac{6-\alpha}{\sqrt{3\beta}}\sin\sqrt{3\beta}\right]^{-1}\nonumber,
\label{fx-Narai}
\end{eqnarray}
where
\begin{equation}
\tilde{f}(x')=\cos^{-1}E+\sqrt{\frac{3\beta}{4}}\left[1-\left(x'\frac{R}{R'}\right)^2\right].
\label{Nariai-1-1}
\end{equation}
The central pressure and sound speed become infinite when $\beta=0.4126$ and the causality limit is $\beta=0.223$.

\subsection*{Quark star like solutions}
In the present  work we  also use  four additional analytical solutions related however to the structure of the called  self-bound stars (i.e. quark stars). In these cases, while the pressure vanishes at the surface, the density remains finite and the solutions are reasonable approximations of strange quark matter stars. Although the density configurations of the mentioned solutions are not suitable to describe the neutron star structure, they are useful for comparison and mainly to examine in which extent the specified configurations affect the main properties of the r-mode. In either  case, valuable information will be obtained.

\subsubsection*{Uniform density}
In the uniform density case (UD) (the Schwarzchild constant-density interior solution), which  has been extensively used in the literature, the density is constant
\begin{equation}
\rho=\frac{3M}{4\pi R^3}={\rm constant}
\label{Un-den-1}
\end{equation}
and the structure function is simple ${\cal F}(x)=1$. Actually, there is no physical justification for this solution since:  a) the energy density does not vanish on the surface of the star  and b) the speed of sound is infinite. Nevertheless, the interiors of dense neutron stars are of nearly uniform density and this solution has some interest~\cite{Schutz-85}.
The solution is applicable for $\beta < 4/9$ otherwise the central pressure becomes infinite.

\subsubsection*{Tolman VI variant (N=1)}
The density distribution is given by~\cite{Postnikov-010}
\begin{equation}
\rho(r)=\frac{3M}{8\pi R^3}\frac{(2-3\beta)(1-3\beta)+\beta(3-7\beta)x^2+2\beta^2x^4}{(1-3\beta+2\beta x^2)^2}.
\label{Tolm-IV-N1}
\end{equation}
%
\subsubsection*{Tolman VI variant (N=2)}
The density distribution function is given by~\cite{Postnikov-010}
\begin{equation}
\rho(r)=\frac{M}{4\pi R^3}\frac{(2-2\beta)^{2/3} (6-15\beta+5\beta x^2)}{(2-5\beta+3\beta x^2)^{5/3}}.
\label{Tolman VI-N1-1}
\end{equation}
%
\subsubsection*{Matese-Whitman I}
The density distribution function is given by~\cite{Matese-80}
\begin{equation}
\rho(r)=\frac{3M}{4\pi R^3}\frac{1-2\beta+2\beta x^2/3}{(1-2\beta+2\beta x^2)^2}.
\label{MW-1}
\end{equation}
%
\subsubsection*{Polytrope $n=1$}

In the present study, the Newtonian  polytropic equation of state $P=K\rho^{1+\frac{1}{n}} $ with $n=1$ has been used for comparison with the realistic solutions of the TOV equations. The density profile is analytical   solution of the corresponding Lane-Emden equation and has the form
\begin{equation}
\rho(r)=\rho_c\frac{\sin\left(\pi x \right)}{\pi x}, \qquad \rho_c=\frac{\pi^2}{3}\frac{3M}{4\pi R^3}.
\label{Polyt-1}
\end{equation}
All the mentioned solutions are functional of the  mass $M$ and  radius $R$ of the neutron star (or the compactness parameter $\beta$ and the central density $\rho_c$). The corresponding equations of state are very general obeying however to the relative mandatory  constraints which ensure that are physical acceptable solutions.

It is worth mentioning  that in a recent study  \cite{Yagi-014} explore the universality of the  I-Love-Q relations. Actually, most of the equations of state used by \cite{Lattimer-01} to establish the universal relation $R=CP^{1/4}$ have been  also used   by \cite{Yagi-014} (and references therein)  to establish the universality of the the I-Love-Q relations. Moreover, \cite{Yagi-014} compared the density profiles correspond to relativistic (TOV equations) and Newtonian  (Lane-Emden equation) treatment of the polytropic equations of state. They found that as relativistic effects become stronger the density profiles become more centrally condensed. They  concluded that although relativistic corrections and rotational corrections do modify the density profiles of stellar configurations, these  modifications are of ${\cal O}$ $(10 \%)$  relative to the results obtained in the Newtonian non-relativistic limit.

Likewise, in a recent work (\cite{Postnikov-010})  the authors calculated the Love numbers $k_2$ by employing both relativistic polytropic equations of state as well as the analytical solutions used in the present work. They found that the Buchdahl and the Tolman VII solutions predict values of $k_2$ that closely track the results for the  $n=1$   polytrope.  The above results lead to the conclusion that the density profiles  are not so sensitive  on the relativistic corrections and could be safely used  to calculate both the time scales and the r-mode instability window.

\section{Results and Discussions}
Firstly, we concentrate our study on the dependence of the  critical angular velocity $\Omega_c$, on the bulk neutron stars properties, that is mass, radius and  density distribution in the case of the fluid neutron star interior. In particular, we  use various analytical solutions of the TOV equations in the framework provided by the  relations~(\ref{Omegac-T-all-1}) and (\ref{Omegac-T-all-2}). Actually, for low values of $T$ the dissipation mechanism is dominated by the shear viscosity and by the bulk viscosity for  high values of $T$.

In Fig.~1(a), we plot the density distribution for the seven selected  analytical solutions,  as well as the corresponding  standard $n=1$ polytropic density profile, that correspond to a neutron star with $M=1.4 M_{\odot}$ and $R=12.53$ Km . The four cases with finite density at the surface are suitable to describe the  quark stars interiors. However, these solutions are taken into account for comparison and also in order to examine with completeness  the density distribution effects on the r-mode instability window. It is also obvious that the Newtonian density profiles of the polytrope $n=1$, which has been used extended in r-mode calculations, deviates from the corresponding relativistic profile mainly at the core of the neutron star. So, it is interesting to examine in which extent this deviation affect the instability window.

In Fig.~1(b), we plot the instability window for the seven analytical solutions mentioned before and the polytropic $n=1$ solution  for $M=1.4 M_{\odot}$ and $R=12.53$ Km. It is obvious that all solutions  predict  similar results. In particular, the three realistic solutions and the polytropic one  lead to a small increase of $\Omega_c$ compared to the other four cases. The use of the polytropic solution, produces almost the same instability window compared to the three relativist solutions. In particular, there is a deviation on the values of $\Omega_c$ less than $4 \%$. It is also concluded that  the instability window exhibits a small dependence on the mass distribution. Actually,  expression~  (\ref{Omegac-T-all-1})  is a  key to explain this behavior. For low values of $T$, that means for ${\cal Y} \ll 1$,   then $\Omega_c \simeq b^{1/6}$. After some algebra we found the relation
\begin{eqnarray}
\Omega_c&\simeq& 362.4 \left(\frac{10^9 {\rm K}}{T}  \right)^{1/3} \left(\frac{10 {\rm km}}{R} \right)^{3/2}
{\cal J}_1^{1/6}, \nonumber \\
 {\cal J}_1&=&\frac{1}{{\cal I}_1^2}
\left[\left(\frac{\rho_c}{10^{16}{\rm gr \ cm^{-3}}}\right)^{1/4}{\cal I}_3^{nn}+1.729{\cal I}_3^{ee}    \right].
\label{Oc-apr-1}
\end{eqnarray}
According to (\ref{Oc-apr-1}), $\Omega_c$ is almost independent  from  the mass $M$, but  depends appreciably on the radius $R$. The factor ${\cal J}_1$ is mainly correlated with the density distribution and depends weakly  on $M$ and $R$. However, due to the exponent $1/6$ the total contribution on $\Omega_c$ is almost negligible. By employing the Tolman VII solution for $M=1.4 M_{\odot}$,  Eq.~(\ref{Oc-apr-1}) takes the form
\begin{equation}
\Omega_c\simeq 706.88 \left(\frac{10^9 {\rm K}}{T}  \right)^{1/3} \left(\frac{10 {\rm km}}{R} \right)^{3/2}.
\label{Oc-apr-1-b}
\end{equation}
For high values of $T$ (${\cal Y} \gg 1$) then $\Omega_c \simeq a^{1/4}$ and we have
\begin{eqnarray}
\Omega_c&\simeq& 13.4  \left(\frac{10 {\rm km}}{R} \right)^{3/4}\left(\frac{M}{M_{\odot}} \right)^{1/2}
\left(\frac{T}{10^9 {\rm K}} \right)^{3/2}{\cal J}_2^{1/4}, \nonumber \\
 {\cal J}_2&=&\frac{{\cal I}_2}{{\cal I}_1^2}.
\label{Oc-apr-2}
\end{eqnarray}
In this case $\Omega_c$ exhibits additional dependence on the mass but the effects of the distribution still remain negligible due to the factor ${\cal J}_2$.

The main conclusion is that $\Omega_c$ mainly depends (for a fixed $T$)  on the neutron star size.  The mass dependence of $\Omega_c$  is more evident  for high $T$. In any case, the effect of the density distribution is negligible.  In the literature     are not only the absolute values of $\Omega_c$  under consideration  but also the ratio $\Omega_c/\Omega_K$. Considering that
\[\Omega_{\rm K}=6650.14 \left(\frac{M}{M_{\odot}}\right)^{1/2}\left(\frac{10 {\rm Km}}{R}\right)^{3/2}\]
the relations~(\ref{Oc-apr-1}) and (\ref{Oc-apr-2}) are rewritten also as
\begin{equation}
\frac{\Omega_c}{\Omega_{\rm K}}\simeq  0.0545 \left(\frac{10^9 {\rm K}}{T}  \right)^{1/3} \left(\frac{M_{\odot}}{M} \right)^{1/2}
{\cal J}_1^{1/6},
\label{Oc-apr-3}
\end{equation}
\begin{equation}
\frac{\Omega_c}{\Omega_{\rm K}}\simeq 0.002 \left(\frac{T}{10^9 {\rm K}}  \right)^{3/2} \left(\frac{R}{10 {\rm km}} \right)^{3/4}
{\cal J}_2^{1/4}
\label{Oc-apr-2-b}
\end{equation}
and the results are presented in Fig.~(2). The mentioned conclusions are displayed also  in Fig.~(3) where we plot $\Omega_c$ for fixed $M=1.4 M_{\odot}$ and various values of the radius, for the Tolman VII solutions (the results by employing the Buchdal and Nariai IV solutions are similar). It is obvious that the  effects of the neutron star size on the instability window are   efficient.

In addition, in Table~1, we present the minimum $T_c^{{\rm min}}$ and maximum $T_c^{{\rm max}}$ critical temperatures (which correspond to the solution of equation $\Omega_c(T)=\Omega_{\rm K}$) as well as  the minimum value of the spin frequency $f_c^{{\rm min}}$ and the corresponding temperature $T_{{\rm min}}$ and ratio $\Omega_c^{{\rm min}}/\Omega_{\rm K}$ for the seven selected analytical solutions (for neutron star with $M=1.4 M_{\odot}$ and $R=12.53$ Km). The values of  $T_c^{{\rm min}}$  are affected by the density distribution. The most realistic distributions (Tolman VII, Buchdahl and Nariai IV) produce higher values of $T_c^{{\rm min}}$. However,   values of  $T_c^{{\rm max}}$ are independent from the mass distribution as  the very strong bulk  dissipation mechanism takes place in  high temperatures. In addition the Tolman VII, Buchdahl and Nariai IV solutions lead to  very similar values of $T_{{\rm min}}$, $f_c^{{\rm min}}$ and $\Omega_c^{{\rm min}}/\Omega_{\rm K}$. The corresponding values for the four quark like solutions are lower.

We also study the effect of the rigid crust on the  r-mode instability window.  In particular, we solve Eq.~(\ref{Omegac-crust-1}) and the results are presented in Fig.~4(a) for the three cases and for fixed mass $M=1.4 M_{\odot}$. Obviously, the effect of density distribution is negligible since all the solutions lead to a similar instability window both for low and high temperatures. In addition, in Fig.~4(b) we present, for the Tolman VII solution that the instability windows correspond to the two cases (rigid crust and without crust) and for three different values of the radius. The effect of the neutron star size is less  pronounced in the crust case. In particular, we found the relation (see below) $\Omega_c \sim R^{-3/2}$ (without crust) and $\Omega_c \sim R^{-4/11}$ (with crust). In any case, Fig~.4(b) confirms previous similar studies related to the strong dissipation mechanism which is active on the crust-core interface  \cite{Wen-012,Haskell-012,Moustakidis-015}.

In order to clarify  further the $\Omega_c-T$ dependence, it is worth  presenting some useful approximations. More precisely,
in a very good approximation, where the viscous dissipation at the boundary layer is stronger  to shear viscosity (that is $\tilde{d}\left(\frac{\Omega_c}{{\rm Hz}}\right)^{1/2} \gg \tilde{b}$) then Eq.~(\ref{Omegac-crust-1}) is written
\begin{equation}
-\left(\frac{\Omega_c}{{\rm Hz}}\right)^{11/2}+\tilde{a}\left(\frac{\Omega_c}{{\rm Hz}}\right)^{3/2}+\tilde{d}=0.
\label{Omegac-crust-2}
\end{equation}
Actually, Eq.~(\ref{Omegac-crust-2}) provides  a very good approximation for all values of $T$. However, it cannot be solved analytically. In any case, it is  interesting to study the temperature dependence of $\Omega_c$ for low values of $T$ which  corresponds to old  and cold neutron stars. In this case, the bulk viscosity mechanism is inactive (that is $\tilde{a}\left(\frac{\Omega_c}{{\rm Hz}}\right)^{3/2} \ll \tilde{d}$) and the desired approximation is written
\begin{equation}
\Omega_c\simeq \left(\tilde{d}\right)^{2/11} \quad ({\rm Hz}).
\label{Omegac-crust-3}
\end{equation}
In order to study further the  $\Omega_c-T$ dependence on bulk neutron star properties we employ the Tolman VII model which, for  $M=1.4 \ M_{\odot}$, is a good approximation for a comprehensive set of realistic equation of states~\cite{Lattimer-01}. In this case, the factor $\tilde{d}$ takes the form
\begin{eqnarray}
\tilde{d}&=&6.125\cdot 10^{18}\left(\frac{10^9 {\rm K}}{T}  \right)  \left(\frac{10 {\rm Km}}{R}\right)^{2}\nonumber\\
 &\times&\left(h(R)^2-\frac{7}{9}h(R)^3\right)^{-2}, \nonumber \\
 && h(R)=1-0.09 \left(\frac{R}{10{\rm km}}\right)^3
\label{tilde-d-3}
\end{eqnarray}
and the critical frequency $\Omega_c$
\begin{eqnarray}
&&\Omega_c\simeq 2605
\left(\frac{10^9 {\rm K}}{T}  \right)^{2/11}  \left(\frac{10 {\rm Km}}{R}\right)^{4/11} {\cal H}(R), \nonumber\\
&&
{\cal H}(R)=\left(h(R)^2-\frac{7}{9}h(R)^3\right)^{-4/11}.
\label{Wc-crust-tol-1}
\end{eqnarray}
To proceed further, and considering that the factor ${\cal H}(R)$  varied very slowly with $R$, we replace it with the mean value ${\cal H}_{{\rm MV}}(R)=1.6736$. This is  a very good approximation for the range $R=10-14$ Km and leads to the simple expression
\begin{equation}
\Omega_c\simeq 4360
\left(\frac{10^9 {\rm K}}{T}  \right)^{2/11}  \left(\frac{10 {\rm Km}}{R}\right)^{4/11} \quad ({\rm Hz}).
\label{Wc-crust-tol-2}
\end{equation}
The above expression is very accurate (the error is less than $0.1\%$) especially in the range $R=12\pm 1$ Km.
In addition, the ratio $\Omega_c/\Omega_K$, for neutron star with mass $M=1.4 \ M_{\odot}$, is given by the expression
\begin{equation}
\frac{\Omega_c}{\Omega_{\rm K}}\simeq 0.554
\left(\frac{10^9 {\rm K}}{T}  \right)^{2/11}  \left(\frac{R }{10 {\rm Km}}\right)^{25/22}.
\label{Wc-crust-tol-3}
\end{equation}
Eq.~(\ref{Wc-crust-tol-2}) could  be used as a measure of the radius. In particular by inverting Eq.~(\ref{Wc-crust-tol-2}) yields
\begin{equation}
R\simeq 1.02\cdot 10^{11}\left(\frac{10^9 {\rm K}}{T}  \right)^{1/2}  \left(\frac{{\rm Hz} }{\Omega_c}\right)^{11/4}  \quad ({\rm Km}).
\label{Wc-crust-tol-4}
\end{equation}
The accurate and  simultaneously measures of $\Omega_c$ and core temperature $T$ may impose constraints on the radius of a neutron star with mass $M=1.4 M_{\odot}$. Additionally, the combination of Eqs.~(\ref{Wc-crust-tol-2}) and (\ref{EOS-7})-(\ref{CnM,1}) and considering that $n_s=0.16 \ {\rm fm}^{-3}$ yields to a directly dependence of $\Omega_c$ on the nuclear equation of state (via the parameter $L$), that is
\begin{equation}
\Omega_c\simeq \left(5794 \pm 108\right)
\left(\frac{10^9 {\rm K}}{T}  \right)^{2/11}   \left(\frac{{\rm MeV}}{L}\right)^{1/11}  ({\rm Hz}).
\label{Wc-crust-tol-6}
\end{equation}
The expression (\ref{Wc-crust-tol-6}) relates the  macroscopic quantity $\Omega_c$ with the microscopic parameter $L$ of the asymmetric nuclear matter in a universal way. The individual characteristic of the EOS is reflected on the uncertainty of the numerical factor in Eq.~(\ref{Wc-crust-tol-6}) as well as on the dependence of the parameter $L$.
The effects of the slope parameter $L$ on the instability window have been recently studied  \cite{Vidana-012,Wen-012,Moustakidis-015}.
The results of the mentioned reference are similar to the predictions of the present work.

We have also  studied the effect of the elasticity of the  crust, via the slippage factor ${\cal S}$, on the instability window. The value ${\cal S}=1$ corresponds to a complete rigid crust without elasticity while lower values of ${\cal S}$ introduce elastic properties to the crust. \cite{Levin-01} showed that the slippage factor is ${\cal S}\approx 0.05-0.1$ in a typical case, while  \cite{Glampedakis-06} found the value ${\cal S}\approx 0.05$.  Following the suggestion of the authors in  Refs.~\cite{Glampedakis-06,Levin-01}, the coefficient $\tilde{d}$ given in Eq.~(\ref{tilde-d-1}) must be multiplied with the factor ${\cal S}^2$. The  approximation (\ref{Wc-crust-tol-2}) is written
\begin{equation}
\Omega_c\simeq 4360 \ {\cal S}^{4/11}
\left(\frac{10^9 {\rm K}}{T}  \right)^{2/11}  \left(\frac{10 {\rm Km}}{R}\right)^{4/11} \quad ({\rm Hz}).
\label{Wc-crust-tol-2-S}
\end{equation}
Obviously, the effect of ${\cal S}$ on the instability window is dramatic, leading to  a large uncertainty on the estimation of $\Omega_c$.
This uncertainty is displayed in Fig.~(5) where the critical spin frequency is plotted for various values of the factor ${\cal S}$ (considering the Tolman VII solution). The effect is more efficient when ${\cal S}$ approaches the typical value ${\cal S}=0.05$. In this case, as expected, the results are similar with the case without crust where the shear viscosity is the dominant  dissipation mechanism at low temperatures. In the same figure, the observed cases of LMXBs and MSRPs from \cite{Haskell-012} are also included for comparison. In particular, we  include many cases of LMXBs and a few of MSRPs (for more details see \cite{Watts-08,Keek-010} and Table~1 of \cite{Haskell-012}). The masses of the mentioned
stars are not measured accurately. In addition, it is worth pointing out that the estimation  of the core temperature has large uncertainty. In any case, it is obvious from Fig.~5 that the location of the stars, inside or outside   the instability window,  depends strongly on the value of the factor ${\cal S}$. In view of the above discussion, it is concluded that the study of the elastic properties of the crust is a very important issue in neutron star physics (for a recent study see \cite{Kobyakov-015}). To clarify further this statement, we display in Fig.~6(a) the instability window, for ${\cal S}=1$ and three values of the slope parameter $L$ (in each case the alike curves correspond to the lower and higher limits). The nuclear symmetry energy effect, via the parameter $L$, is weakly. In particular, by combining Eqs.~(\ref{Wc-crust-tol-6}) and (\ref{Wc-crust-tol-2-S}) we find
\begin{eqnarray}
\Omega_c&\simeq& \left(5794 \pm 108\right)\ {\cal S}^{4/11}\nonumber\\
&\times&
\left(\frac{10^9 {\rm K}}{T}  \right)^{2/11}   \left(\frac{{\rm MeV}}{L}\right)^{1/11} \quad ({\rm Hz}).
\label{Wc-crust-tol-7}
\end{eqnarray}
Considering that the value of $L$ increases from  $20$ MeV to $110$ MeV  then the value of  $\Omega_c$ decreases around 17$\%$.
However, as displayed in Fig.~6(b), when  the slippage factor is taken into account, its effect is even stronger compared to the effect of $L$. Consequently, it is very important, in order to provide reliable information for the instability window, to employ  accurate values for the factor ${\cal S}$.
Eq.~(\ref{Wc-crust-tol-2-S}) offers another possibility,  the rough estimation of $S$ via the expression
\begin{equation}
{\cal S}\simeq 9.8 \cdot 10^{-11} \left(  \frac{\Omega_c}{{\rm Hz}}  \right)^{11/4}
\left(\frac{T}{10^9 {\rm K}}  \right)^{1/2}  \left(\frac{R}{10 {\rm Km}}\right).
\label{Wc-crust-tol-2-inv}
\end{equation}
It is obvious that  accurate and simultaneous measures of $\Omega_c$, $R$ and $T$ may impose constraints on  ${\cal S}$ and consequently  on the measure of the crust elasticity. The knowledge of ${\cal S}$ will provide important information on the crust  structure. Moreover, the measure of ${\cal S}$ will be used as a  useful tool to check relevant theoretical predictions (\cite{Levin-01,Kinney-03,Glampedakis-06} and references therein). In any case, the rigidity  of the crust appears to  be the most efficient damping mechanism. However, much more work is necessary in order to clarify further this issue. A reliable theoretical calculation of the slippage factor ${\cal S}$ in comparison with observation measures, may  reveal the magnitude of crust elasticity  and provide more useful insights to this open problem.

Finally, we studied the effects of the mutual friction on the instability window in comparison to the minimal model and the crust viscosity effects.
The corresponding  time scale $\tilde{\tau}_{_{MF}}$ varying in the large range $5 \ {\rm s} \leq  \tilde{\tau}_{_{MF}} \leq 10^4\ {\rm s} $ in order to systematically study the mutual friction effects (see  Fig.~7).  We confirm the results of the previous work of   \cite{Lidblom-2000b} where the MF effects are almost negligible for $\tilde{\tau}_{_{MF}} >50 \ {\rm s}$.  In this case, the main viscosity mechanism is due to the Ekman layer viscosity and the previous analysis concerning the r-mode from the equation of state is a good approximation. However, for $\tilde{\tau}_{_{MF}} < 50 \ {\rm s}$ the mutual friction effects are very important narrowing remarkably the instability window. In particular, for $\tilde{\tau}_{_{MF}} \simeq 5 \ {\rm s}$ the window disappears that is the mutual friction suppresses completely the gravitational radiation. In this case, since the mutual friction suppression overcomes significantly those due to the Ekman layer the value of the time scale  $\tilde{\tau}_{_{MF}}$ is the dominate factor and further analysis is essential in order to clarify further the role of the equation of state. Actually, in this case and in a good approximation, the equilibrium equation takes the simple form $t_{_{MF}}=|t_{_{GR}}|$ and the critical angular velocity $\Omega_c$, for the Tolman VII solution,  is given by
\begin{equation}
\Omega_c=43.2 \frac{1}{\tilde{\tau}_{_{MF}}}\frac{1}{\beta^{7/2}}.
\label{Omegac-mf-gr}
\end{equation}
It is obvious that, in this special case, the $\Omega_c$ is very sensitive on the compactness parameter $\beta$. The most compact configuration of a neutron star leads to dramatic lowering of  the critical angular velocity values. For example when the value of $\beta$ varies on the interval $0.1-0.2$ then (and for a the typical value  $\tilde{\tau}_{_{MF}}=8\ {\rm s}$) the $\Omega_c$ varies on the large interval $17076-1510 \ {\rm Hz}$. In addition, the combination of Eqs.~(\ref{Omegac-mf-gr}) and (\ref{EOS-7})-(\ref{CnM,1}) and considering that $M=1.4 \ M_{\odot}$ yields to a dependence of $\Omega_c$ on  the parameter $L$, that is
\begin{equation}
\Omega_c\simeq \left(2757 \pm 492\right)
\frac{1}{\tilde{\tau}_{_{MF}}} \left(\frac{L}{{\rm MeV}}\right)^{7/8}  ({\rm Hz}).
\label{Wc-L-MF}
\end{equation}
In any case, it is worth pointing that according to the analysis of \cite{Lidblom-2000b} only $2 \%$  of the expected range of $\epsilon$ leads to the time scale  $\tilde{\tau}_{_{MF}}$ shorter than $15 \ {\rm s} $ in neutron stars with temperature  about $10^8\ {\rm K}$ (that are typical for low mass x-rays binaries).

In a recent paper, \cite{Haskell-09} studied more in details the extent in which  the mutual friction can further restrict the range where the r-mode is unstable. Their study has the advantage of using realistic pairing gaps and allowing the mutual friction parameters to take the whole range of permissible values. Those they considered both the weak and the strong drag regime.
In any case they concluded, in accordance with the previous work of \cite{Lidblom-2000b} and \cite{Lee-03},  that in the weak drag regime the mutual friction is not the leading damping mechanism for the r-mode. Consequently, and according to the mentioned papers,  we expect that for low temperature the Ekman layer is the dominant dissipation mechanism. This statement, also amplifies our estimation that, at least at low temperature, the minima model+crust is a good approximation for the description and study  of the instability window.

In a recent work, \cite{Alford-2014} demonstrated that that precise pulsar timing data can constrain the star's composition, through unstable global oscillations whose damping is determined by microscopic properties of the interior. They studied  both the standard static instability boundaries as well as the dynamical  instability boundaries.
They employed the minimal hadronic model where only viscous damping has been included. Various others damping mechanism as superfluid pairing (including the mutual friction), magnetic field, hyperonic matter etc. had  not been  taken into account.
In particular they shown that ungapped interacting quark matter is consistent with both the observed radio and x-ray data, whereas for ordinary nuclear matter some additional enhanced damping mechanism is required.

In comparison, in the present work we concentrate our study on the static instability window paying special attention on the crust effect as well as on the damping effects of the mutual friction. We verified previous studies  that the mutual effects are important and under some assumptions  could explain the observation data, concerning old cold neutron star,  even in the case of hadronic matter. Actually, in the special case  when the effects of the mutual friction are very strong, the critical angular velocity $\Omega_c$ depends strongly on the compactness parameter $\beta$ as well on the slope parameter $L$.
In any case, as discussed above,    additional theoretical study is required in order to establish in details  the mutual friction effects on the instability window.

Here, it is also worth pointing out  that in the present work the r-mode has been studied in the framework of the Newtonian theory of oscillations     but with the use of analytical solutions of the relativistic  TOV equations. However, the most proper and accurate procedure is the relativistic formulation of the r-mode and the use of realistic  equation of states (or analytical solution of the TOV equations), where the viscosity must be taken into account in a self-consistent way. In a  recent paper, \cite{Idrisy-015}  exhibited  the role of the relativistic corrections. In particular, they found that for
realistic equations of state,  the r-mode frequency ranges from $1.39-1.57$ times the spin frequency of the star when the relativistic compactness parameter $\beta$ is varied over the astrophysically motivated interval $0.11-0.31$. In any case, a further theoretical work must dedicated to the mentioned formalism, in order to clarify further the r-mode oscillation problem  both in  slow  and rapid rotating neutron stars.

It should be noted that in the present analysis additional degrees of freedom, like quarks and hyperon matter as well as  the strong magnetic field are  not considered. It is well known that the presence of quark and hyperons influences the dissipation mechanisms
since one has to take into account the shear and also the bulk
viscosities due to the presence of this kind of matter. Actually,
there are several recent studies in this
direction~\cite{Alford-2012b,Haskell-012,Andersson-010,Haskell-07,Linddblom-02,Rupak-013,Chatterjee07,Gusakov-013}.
In any case, when more degrees of freedom are taken into account,
the analysis and the prediction of the related instability window  become more complete and consequently  more reliable.

\section{Summary and Conclusions}
In the present work we investigated  r-mode constraints from the neutron star equation of state. Firstly,  we examined the case of a neutron star with a fluid interior and we derived an analytical solution for the $\Omega_c-T$ dependence. In particular, we  used  a set of analytical solution of the TOV equations in order to  reveal the role of the bulk neutron star properties (radius, mass, mass distribution) on the r-mode instability window. The main findings include the strong dependence of $\Omega_c$ on the neutron star size and the very  weakly dependence on the other two properties for low values of temperature.  Secondly,  we examined the more realistic case where the effect of the solid crust is included in our study. In this case we found that  the effect of the radius is also the most important but the dependence is more weakly compared to the fluid interior case. In any case, the dissipation effect due to the solid crust decreases considerably the instability window.

In view of the above results and motivated by the strong radius dependence of the critical angular velocity, we propose a phenomenological approach in order to correlate $\Omega_c$ with microscopic properties of the nuclear equation of state. This approach, despite of its simplicity,  provides a few insights on the study of the effects of the EOS on the r-mode instability window, in a universal way. In particular, the radius of a NS depends strongly on the specific character of the EOS for densities close to the saturation density. By employing an  empirical  relation, we related the $\Omega_c$ to the slope parameter $L$ which is an individual characteristic of any EOS. We also  proposed  an approximated formula for the $\Omega_c-L$ dependence applicable for a large number of EOS.
This approach leads to some simplified empirical  relations. Moreover, the proposed method provides, in a way,  constraints on the nuclear equation of state with the help of accurate measurements of the main bulk neutron star properties.
We also examined   the case of an elastic crust via the slippage factor ${\cal S}$. We found that this factor is the most important, concerning the estimation of the instability window. The measure of ${\cal S}$ is of importance, in order to define reliable estimation of the corresponding instability window. On the other hand, we proposed possible measure of ${\cal S}$ in the case of accurate measures of $\Omega_c$,  $R$ and $T$.

Finally, we verified previous studies  that the mutual effects are very important and under some assumptions  could explain the observation data, concerning old cold neutron star,  even in the case of hadronic matter. However, more theoretical work is appropriate in order to establish in details the mutual friction dissipations effects and  to clarify further the equation of state constraints on the r-mode instability window.

\section*{Acknowledgments}
This work was supported by  the Aristotle University of Thessaloniki Research Committee under Contract No. 89286.

\begin{figure*}
\centering
\includegraphics[height=8.5cm,width=8.5cm]{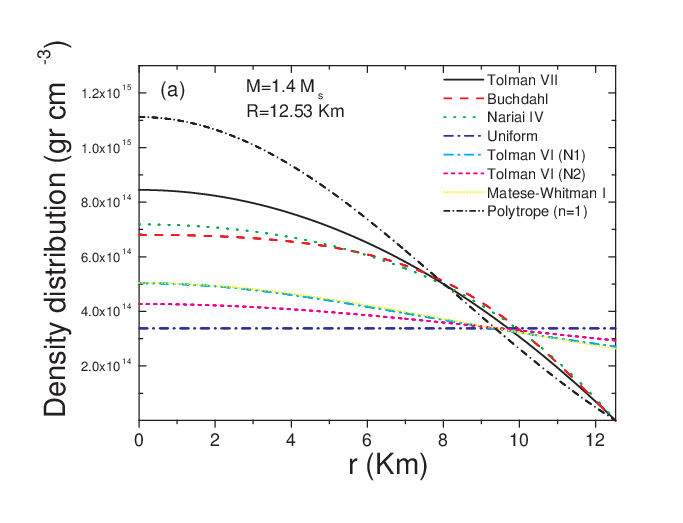}
\includegraphics[height=8.5cm,width=8.5cm]{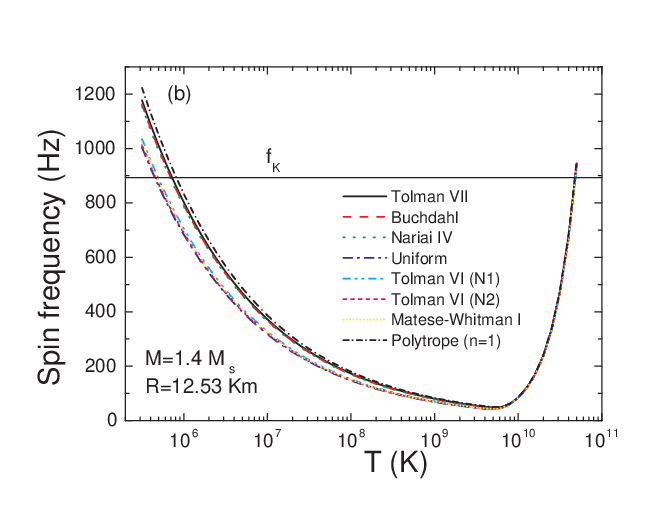}
\caption{ (a) The density distribution $\rho(r)$,  for the seven
selected  analytical solutions of the TOV equations and the standard $n=1$ polytropic density profile,  for neutron
star with $M=1.4 \ M_{\odot}$ and radius $R=12.53$ Km. (b)  The instability window,  for the seven
selected  analytical solutions of the TOV equations and the standard $n=1$ polytropic density profile,  for neutron
star $M=1.4 \ M_{\odot}$ and radius $R=12.53$ Km. The
thin solid line corresponds to the Kepler frequency ${\rm f}_{\rm
K}=893\ {\rm Hz}$.   }
 \label{}
\end{figure*}


\begin{figure}
\vspace{0cm}
\centering
\includegraphics[height=8.5cm,width=8.0cm]{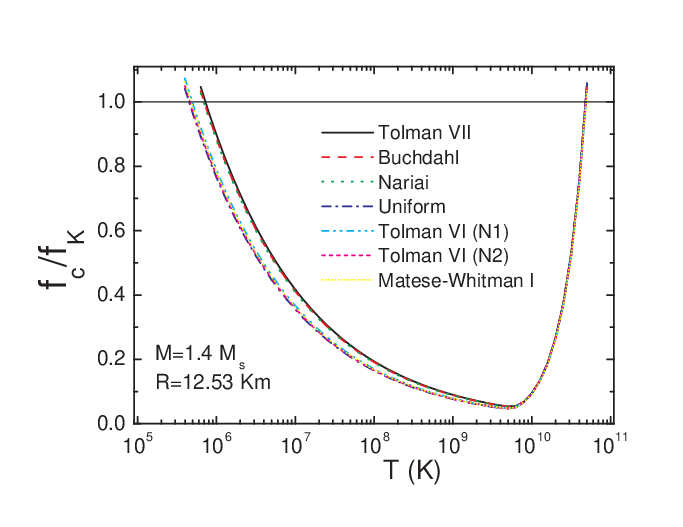}
\vspace{0.0cm}
\caption{ The ratio $f_c/f_{\rm K}$ as a
function of the temperature for the
seven selected  analytical solution.  } \label{}
\vspace{0.0cm}
\end{figure}
\begin{figure}
\centering
\includegraphics[height=8.5cm,width=8.0cm]{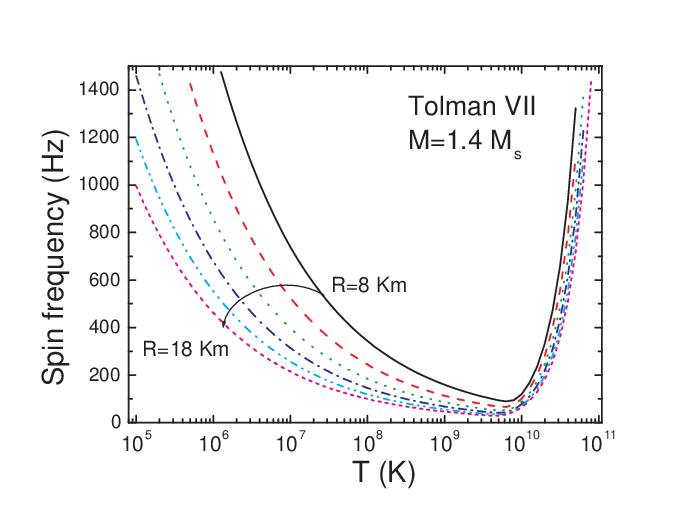}
\vspace{0cm} \caption{The instability window which corresponds to the Tolman VII
solution for $M=1.4 \ M_{\odot}$ and various values of the radius
$R$. } \label{}
\end{figure}

\begin{figure}
\centering
\includegraphics[height=8.5cm,width=8.0cm]{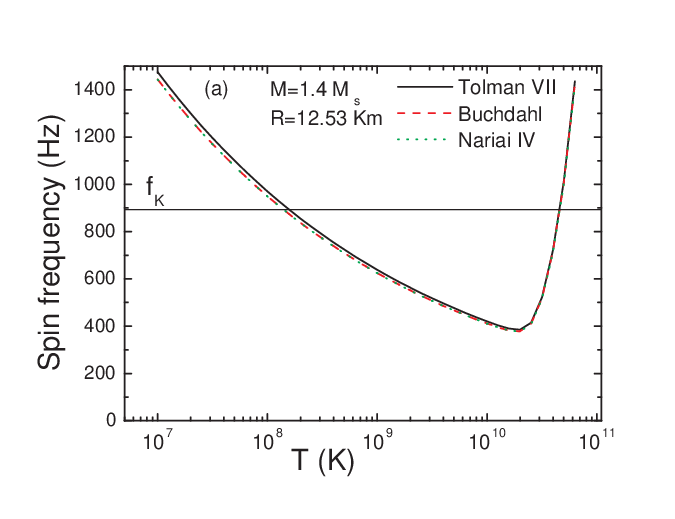}
\includegraphics[height=8.5cm,width=8.0cm]{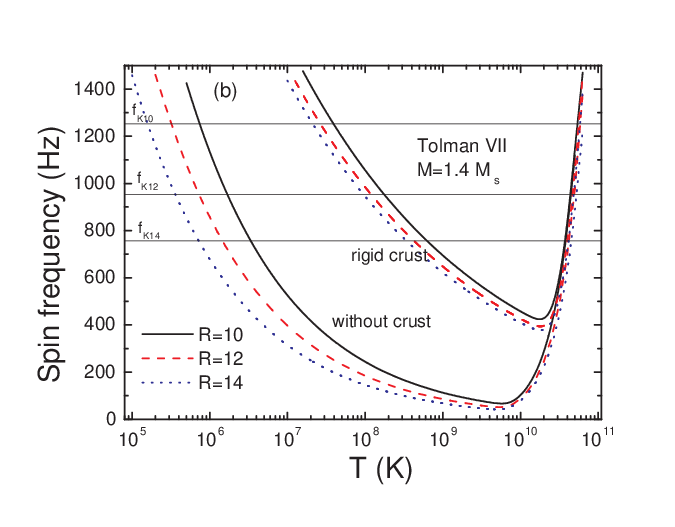}
\vspace{0cm} \caption{(a) The instability window for the Tolman VII, Buchdahl and Nariai IV
solutions when the effect of the crust has
been included. (b) The instability window for the Tolman VII
solution, with and without crust effects, for three different
values of $R$. The corresponding Kepler frequencies have been also  included
 for comparison.  } \label{}
\end{figure}
\begin{figure}
\centering
\includegraphics[height=8.5cm,width=8.0cm]{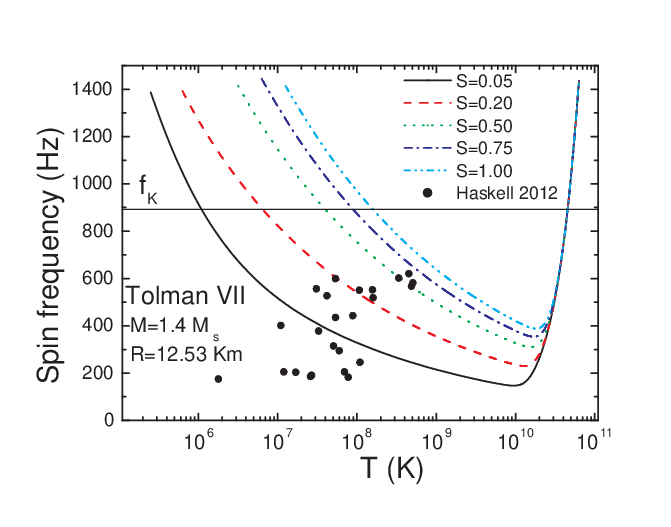}
\vspace{0cm} \caption{The instability window for the Tolman VII
solution when the elasticity of the crust is taken into account
via the slippage factor ${\cal S}$. The observed cases of LMXBs and MSRPs
from \cite{Haskell-012} are also included for
comparison.  } \label{}
\end{figure}

\begin{figure}
\centering
\includegraphics[height=8.5cm,width=8.0cm]{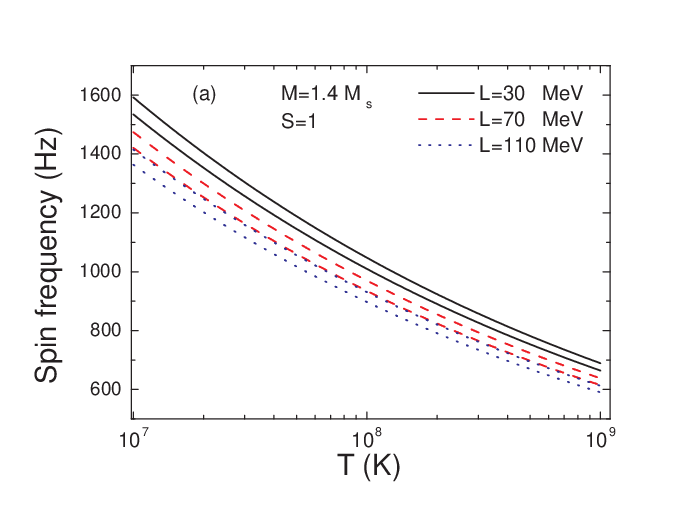}
\includegraphics[height=8.5cm,width=8.0cm]{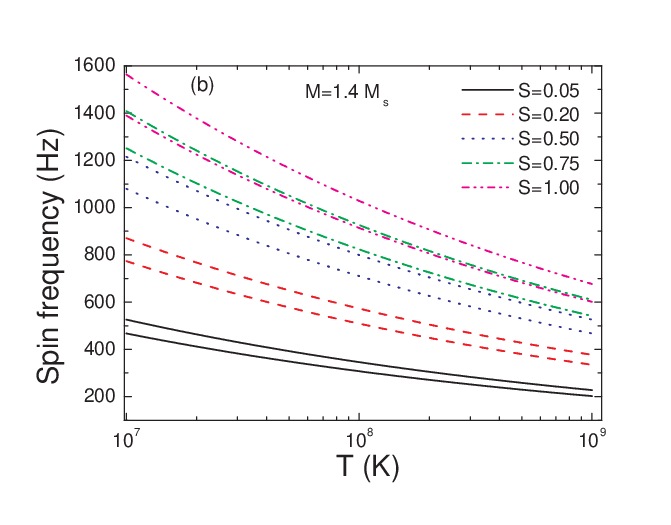}
\vspace{0cm} \caption{(a) The instability window for rigid crust (slippage factor ${\cal S}=1$) and for various values of the slope parameter $L$. In each case, the range of the bands is incorporated  due to the uncertainties  via the values of $C[n_s,M=1.4 M_{\odot}]$ (see Eq.~(\ref{Wc-crust-tol-6}). (b) The instability window for various values of the factor ${\cal S}$. In each case the range of the bands corresponds to $L=30 $ MeV (upper curve) and $L=110$ MeV (lower curve) (see Eq.~(\ref{Wc-crust-tol-7})). The uncertainties due to the values of $C[n_s,M=1.4 M_{\odot}]$ are averaged. } \label{}
\end{figure}
\begin{figure}
\centering
\includegraphics[height=8.5cm,width=8.0cm]{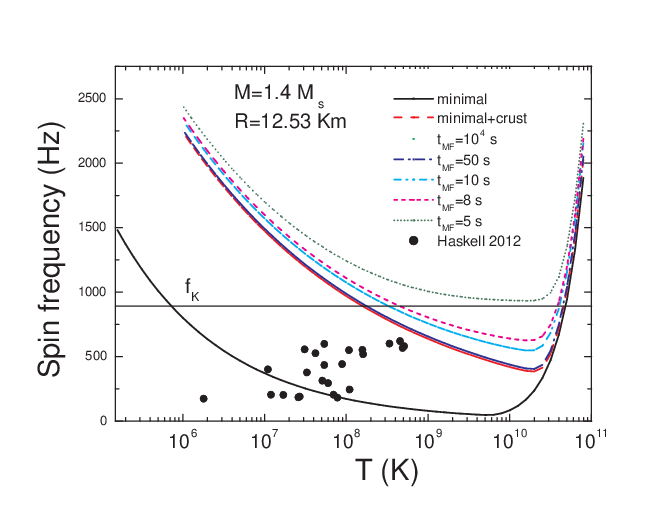}
\vspace{0cm} \caption{The instability window for the Tolman VII
solution  for the cases a)  minimal model, b) minimal model+crust considering slippage factor ${\cal S}=1$ and c) minimal model+crust including also mutual friction effects for various values  of the  time scale $\tilde{\tau}_{_{MF}}$. The observed cases of LMXBs and MSRPs
from \cite{Haskell-012} are also included for
comparison.
} \label{}
\end{figure}

 \begin{table*}
 \small
\caption{The minimum $T_c^{{\rm min}}$ and maximum $T_c^{{\rm max}}$ critical temperatures (which correspond to the solutions of the equations $\Omega_c(T)=\Omega_{\rm K}$), the minimum value of the spin frequency $f_c^{{\rm min}}$ and the corresponding temperature $T_{{\rm min}}$ and ratio $\Omega_c^{{\rm min}}/\Omega_{\rm K}$ for the selected analytical solutions. All the values correspond to the case of the fluid interior of neutron stars. }
 \label{t:1}
\vspace{0.5cm}
\begin{tabular}{@{}|cccccc|@{}}
\hline
 {\rm Models}     &   $T_c^{{\rm min}}(\times 10^5)$ (K)    & $T_c^{{\rm max}}(\times 10^{10})$(K)   &  $T_{{\rm min}}(\times 10^9)$ (K)   &   $f_c^{{\rm min}}$ (Hz) &  $\Omega_c^{{\rm min}}/\Omega_{\rm K}$        \\
\hline
\hline
 {\rm Tolman VII}            & 7.24    & 4.83     &     5.44           &      48              &   0.054        \\
 \hline
{\rm Buchdahl}               &  6.97    &  4.82       &   5.40             &    47                  &  0.053         \\
 \hline
  {\rm Narai IV}             &  6.83    &  4.83   &     5.39           &         47             &    0.053        \\
 \hline
 {\rm Uniform}               &  4.49     &  4.83      &    4.99            &     42                 &   0.047          \\
 \hline
 {\rm Tolman IV (N=1)}       &  4.91   &   4.87   &     5.11           &         42            &  0.048       \\
 \hline
 {\rm Tolman IV (N=2)}       & 4.60     &  4.86   &     5.04           &        42              &   0.047        \\
 \hline
 {\rm Matese-Whitman I}     & 4.76     &  4.88   &      5.09          &         42             &  0.048 \\
 \hline
\end{tabular}
\end{table*}


 \begin{table*}
 \small
\caption{ The same with the Table I when the effects of the crust have been included.  }
 \label{t:1}
\vspace{0.5cm}
\begin{tabular}{@{}|cccccc|@{}}
\hline
 {\rm Models}     &   $T_c^{{\rm min}}(\times 10^8)$ (K)    & $T_c^{{\rm max}}(\times 10^{10})$(K)   &  $T_{{\rm min}}(\times 10^{10})$ (K)   &   $f_c^{{\rm min}}$ (Hz) &  $\Omega_c^{{\rm min}}/\Omega_{\rm K}$        \\
\hline
\hline
 {\rm Tolman VII}            & 1.58    & 4.60     &   1.90             &   385                 &   0.431        \\
 \hline
{\rm Buchdahl}               & 1.41    &  4.62       &  1.88             &     377                 &  0.422         \\
 \hline
  {\rm Narai IV}             & 1.40     &  4.64   &    1.89            &     377                 &     0.421       \\
 \hline
 \end{tabular}
\end{table*}





\begin{thebibliography}{}
%
\bibitem{Adler-74} Adler, R.J.,  J. Math. Phys. {\bf 15},  727 (1974)
%
\bibitem{Alford-2012b} Alford, M.G.,   Mahmoodifar, S.,  and
Schwenzer, K.,  Phys. Rev. D {\bf 85}, 044051 (2012)
%
\bibitem{Alford-2014} Alford, M.G., and
Schwenzer, K., Phys. Rev. Lett., {\bf 113}, 251102 (2014)
%
\bibitem{Andersson-1998} Andersson, N., Astrophys. J. {\bf 502}, 708 (1998)
%
\bibitem{Andersson-2003} Andersson, N.,  Class. Quantum Grav. {\bf
20}, R105 (2003)
%
\bibitem{Andersson-2001} Andersson, N.,  and  Kokkotas, K.D.,  Int. J.
Mod. Phys. D {\bf 10}, 381 (2001)
%
\bibitem{Andersson-99} Andersson, N.,  Kokkotas, K.D.,  and Schutz, B.F.,
Astrophys. J. {\bf 510}, 846 (1999)
%
\bibitem{Andersson-2000} Andersson, N.,   Jones, D.I.,  Kokkotas, K.D.,
and  Stergioulas, N.,  Astrophys. J. {\bf 534}, L75 (2000)
%
\bibitem{Andersson-010} Andersson, N.,   Haskell, B.,  and  Comer, G.L.,  Phys. Rev. D {\bf 82}, 023007 (2010)
%
\bibitem{Bildsten-2000} Bildsten, L., and Ushomirsky, G.,  Astrophys. J.
Lett. {\bf 529}, L33 (2000)
%
\bibitem{Bondarescu-09} Bondarescu, R.,   Teukolsky, S.A.,  and
Wasserman, I.,  Phys. Rev. D {\bf 79}, 10403 (2009)
\bibitem{Buchdahl-67} Buchdahl, H.A.,  Astrophys. J. {\bf 147}, 310 (1967)
\bibitem{Chirenti-2015}  Chirenti, C.,   de Souza, G.H.,  and  Kastaun, W.,  Phys. Rev. D {\bf 91}, 044034 (2015)
%
\bibitem{Chatterjee07}  Chatterjee, D.,  and  Bandyopadhyay, D.,  Astrophys. Space Sci. {\bf 308}, 451 (2007)
\bibitem{Delgaty-98} Delgaty, M.S.R.,  and  Lake, K.,  Comput. Phys. Commun. {\bf 115}, 395 (1998)
\bibitem{Demorest-010} Demorest, P.B.,   Pennucci, R.,   Ransom, S.M.,
Roberts, M.S.E.,  and  Hessels, J.W.T.,  Nature {\bf 467}, 1081 (2010)
\bibitem{Fattoyev-014} Fattoyev, F.J.,    Newton, W.G., and Li, B.A., Eur. Phys. Jour. A {\bf 50}, 45 (2014)
%
\bibitem{Friedman-98}Friedman, J.L.,  and  Morsink, S.M.,  Astrophys. J. {\bf 502}, 714 (1998)
%
\bibitem{Friedman-99}Friedman J.L.,  and Lockitch, K.H., Prog. Theor.
Phys. Suppl. {\bf 136}, 121 (1999)
%
\bibitem{Glampedakis-06} Glampedakis, K.,  and  Andersson, N.,   Phys. Rev. D {\bf 74}, 044040 (2006)
%
\bibitem{Glendenning-2000} Glendenning, N.K.,  {\it Compact Stars:
Nuclear Physics, Particle Physics, and General Relativity}
(Springer, Berlin, 2000)
%
\bibitem{Gusakov-014} Gusakov, M.E.,  Chugunov, A.I.,  and  Kantor, E.M.,  Phys. Rev. D {\bf 90}, 063001 (2014)
%
\bibitem{Gusakov-013} Gusakov, M.E.,   Kantor, E.M.,   Chugunov, A.I.,  and Gualtieri, L.,  MNRAS {\bf 428}, 1518 (2013)
%
\bibitem{Haensel-07} Haensel, P.,  Potekhin, A.Y.,  and Yakovlev, D.G.,
{\it Neutron Stars 1: Equation of State and Structure}
(Springer-Verlag, New York, 2007)
\bibitem{Haskell-15}Haskell, B., Int.J. Mod. Phys. E {\bf 24}, 1541007 (2015).

\bibitem{Haskell-012}  Haskell, B.,   Degenaar, N.,   Ho, W.C.G.,
Mon. Not. R. Astron, Soc. {\bf 424}, 93 (2012)
%
\bibitem{Haskell-07} Haskell, B.,   Andersson, N.,   Jones, D.L.,  and  Samuelsson, L.,  Phys.
Rev. Lett., {\bf 99}, 1101 (2007)
%
\bibitem{Haskell-09} Haskell, B.,   Andersson, N., Passamonti, A.,   Mon. Not. R. Astron. Soc. {\bf 397}, 1464 (2009)
%
\bibitem{Ho-011} Ho, W.C.G.,  Andersson, N.,  and  Haskell, B., Phys. Rev. Lett., {\bf 107}, 101101 (2011)
%
\bibitem{Idrisy-015} Idrisy, A., Owen, B.J.,  and Jones, D.I., Phys. Rev. D {\bf 91}, 024001 (2015)
%
\bibitem{Keek-010} Keek, L.,   Galloway, D.K.,  in't Zand, J.J.M.,  and
 Heger, A.,  Astrophys. J. {\bf 718}, 292 (2010)
\bibitem{Kinney-03} Kinney, J.,  and  Mendell, G.,  Phys. Rev. D {\bf 67}, 024032 (2003)
%
\bibitem{Kobyakov-015} Kobyakov, D.,  and  Pethick, C.J.,  MNRAS {\bf 449}, L110, (2015)
%
\bibitem{Kokkotas-99} Kokkotas, K.D.,  and  Stergioulas, N.,  Astron.
Astrophys. {\bf 341}, 110 (1999)
%
\bibitem{Kokkotas-015} Kokkotas, K.D.,  and  Schwenzer, K., arXiv:1510.07051[gr-qc]
%
\bibitem{Kolomeitsev-014}  Kolomeitsev, E.E.,  and  Voskresensky, D.N.,  Eur.Phys. J. A {\bf 50}, 180 (2014)
%

\bibitem{Lattimer-012} Lattimer, J.M.,   Annu. Rev. Nucl. Part. Sci. {\bf 62}, 485 (2012)
%
\bibitem{Lattimer-01} Lattimer, J.M.,  and  Prakash, M.,  Astrophys. J. {\bf 550}, 426 (2001)
\bibitem{Lattimer-07}  Lattimer, J.M.,  and M. Prakash, M.,  Phys. Rep.
{\bf 442}, 109 (2007)
%
\bibitem{Lattimer-013}  Lattimer, J.M., and  Lim, Y.,  Astrophys. J. {\bf 771}, 51 (2013)
%

\bibitem{Lattimer-05} Lattimer, J.M.,  Neutron Stars, lectures delivered at the 33rd Summer Institute on Particle Physics, SSI 2005, (unpublished)
%
\bibitem{Lattimer-2000} Lattimer, J.M.,  and  Prakash, M.,  Phys. Rep.  {\bf 333-334}, 121 (2000)
%
\bibitem{Lattimer-05b} Lattimer, J.M.,  and  Prakash, M.,  Phys. Rev. Lett., {\bf 94}, 1105 (2005)
%
\bibitem{Lee-03}  Lee, U., Yoshida, S.,   Astrophys. J. {\bf 586}, 403 (2003)
%
\bibitem{Levin-01} Levin, Y.,  and  Ushomirsky, G.,  Mon. Not. R. Astron. Soc. {\bf 324}, 917 (2001)
%
\bibitem{Lidblom-2000a}Lindblom,L.,  Owen, B.J., and Ushomirsky, G., Phys. Rev. D {\bf 62}, 084030 (2000a)
\bibitem{Lidblom-2000b}Lindblom, L.,   and Mendell, G., Phys. Rev. D {\bf 61}, 104003 (2000b)

\bibitem{Linddblom-02} Lindblom, L.,  and Owen, B.J.,  Phys. Rev. D {\bf 65}, 063006 (2002)
%
\bibitem{Lidblom-98}  Lindblom, L.,   Owen, B.J.,  and  Morsink, S.M.,  Phys.
Rev. Lett., {\bf 80}, 4843 (1998)
%
\bibitem{Mahmoodifar-013} Mahmoodifar, S.,  and  Strohmayer, T.,  Astrophys. J. {\bf 773}, 140 (2013)
%
\bibitem{Matese-80} Matese, J.,  and  Whitman, M.,  Phys. Rev. D {\bf 22}, 1270 (1980)
%
\bibitem{Moustakidis-015} Moustakidis, Ch.C.,  Phys. Rev. C {\bf 91}, 035804 (2015)
\bibitem{Mytidis-015} Mytidis, A.,  Coughlin, M.,  and  Whiting, B.,   Astrophys. J. {\bf 810}, 27 (2015)
%
\bibitem{Nariai-50} Nariai, H.,  Sci. Rep. Tohoku Univ. Ser. 1 {\bf 34}, 160 (1950)
\bibitem{Nariai-51} Nariai, H.,  Sci. Rep. Tohoku Univ. Ser. 1 {\bf 35}, 62 (1951)
\bibitem{Nariai-99} Nariai, H.,  Gen. Rel. and Grav. {\bf 31}, 951 (1999)
\bibitem{Newton-014}    Newton, W.G., Hooker, J., Gearheart, M., Murphy, K., Wen, D.H., Fattoyev, F.J., and Li, B.A., Eur. Phys. Jour. A {\bf 50}, 41 (2014)
%
\bibitem{Owen-98}Owen, B.J.,  Lindblom, L.,  Cutler, C.,  Schutz, B.F.,  Vecchio, A.,  and Andersson, N.,
Phys. Rev. D {\bf 58}, 084020 (1998)
%
\bibitem{Prakash-014-a} Prakash, M., Nucl. Phys. {\bf A 928}, 260 (2014)
\bibitem{Prakash-94}  Prakash, M.,  The Equation of State and Neutron
Stars, lectures delivered at the Winter School held in Puri India,
1994 (unpublished)
%
\bibitem{Postnikov-010} Postnikov, S.,   Prakash, M.,   Lattimer, J.M.,  Phys. Rev. D {\bf 82}, 024016 (2010)
%
\bibitem{Raghoonundun-015} Raghoonundun, A.M., and Hobill, D.W.,    Phys. Rev. D {\bf 92}, 124005 (2015)

%
\bibitem{Read-09-a} Read,  J.S.,  Markakis, C.,   Shibata, M.,    Uryu, K.,  Creighton, J.D.E.,  and  Friedman, J.L.,  Phys. Rev. D {\bf 79}, 0124033 (2009)
%
\bibitem{Read-09-b} Read, J.S.,   Lackey, B.D.,   Owen, B.J.,  Friedman, J.L.,  Phys. Rev. D {\bf 79}, 0124032 (2009)
%
\bibitem{Rieutord-2000} Rieutord, M.,  astro-ph/0003171
%
\bibitem{Rupak-013} Rupak, G.,  and  Jaikumar, P.,  Phys. Rev. C {\bf 88}, 065801 (2013)
%
\bibitem{Schutz-85} Schutz, B.F.,  {\it  A First Course in General Relativity}, (Cambridge University Press, Cambridge, 1985)
%
\bibitem{Shapiro-83}Shapiro, S.L., and Teukolsky, S.A.,  {\it Black
Holes, White Dwarfs, and Neutron Stars} (John Wiley and Sons, New
York, 1983)
\bibitem{Rezzolla-014} Takami, K.,   Rezzolla, L.,  and  Baiotti, L.,   Phys. Rev. Lett., {\bf 113}, 091104 (2014)
%
\bibitem{Tolman-39} Tolman, R.C.,  Phys. Rev. {\bf 55}, 364 (1939)
%
\bibitem{Vidana-012} Vida\~{n}a, I.,  Phys. Rev. C {\bf 85}, 045808 (2012); Erratum  Phys. Rev. C {\bf 90}, 029901 (2014)
%
\bibitem{Watts-08} Watts, A.L.,   Krishnam, B.,  L. Bildsten, L.,  and
Schutz, B.F.,  Mon. Not. R. Astron. Soc. {\bf 389}, 839 (2008)
%
\bibitem{Watts-016} Watts, A.L., et al., arXiv: 1602.01081 [astro-ph.HE]
%
\bibitem{Wen-012} Wen, D.H.,  Newton, W.G.,  and  Li, B.A.,  Phys. Rev. C {\bf 85}, 025801 (2012)
%
\bibitem{Yagi-014} Yagi, K., Stein, L.C., Pappas, G., Yunes, N., and Apostolatos, T.,  Phys. Rev. D {\bf 90}, 063010 (2014)



\end{thebibliography}
\end{document}